\documentclass[aps,prb,twocolumn,superscriptaddress]{revtex4-1}

\usepackage{graphicx}
\usepackage{dcolumn}
\usepackage{bm}
\usepackage{epstopdf}
\usepackage{float}
\usepackage{amsmath}
\usepackage{color}
\usepackage{comment}
\usepackage[dvipsnames]{xcolor}

\usepackage[colorlinks=true,
            linkcolor=blue,
            citecolor=blue,
            urlcolor=black,
            filecolor=black
           ]{hyperref}

\begin{document}
\title{Probing  spatially resolved spin density correlations with trapped excitons}
\author{Shanshan Ding}
\affiliation{Center for Complex Quantum Systems, Department of Physics and Astronomy, Aarhus University, Ny Munkegade, DK-8000 Aarhus C, Denmark}
\affiliation{Institute of Atomic and Molecular Physics, Sichuan University, Chengdu 610065, China}
\author{Jose Antonio Valerrama Botia}
\affiliation{Center for Complex Quantum Systems, Department of Physics and Astronomy, Aarhus University, Ny Munkegade, DK-8000 Aarhus C, Denmark}
\affiliation{Quantum Science Center of Guangdong-Hong Kong-Macao Greater Bay Area (Guangdong), Shenzhen 508045, China}
\author{Aleksi Julku}
\affiliation{Center for Complex Quantum Systems, Department of Physics and Astronomy, Aarhus University, Ny Munkegade, DK-8000 Aarhus C, Denmark}
\author{Zhigang Wu}
\affiliation{Quantum Science Center of Guangdong-Hong Kong-Macao Greater Bay Area (Guangdong), Shenzhen 508045, China}
\author{G. M. Bruun}
\affiliation{Department of Physics and Astronomy, Aarhus University, Ny Munkegade, DK-8000 Aarhus C, Denmark}

\date{\today}

\begin{abstract}
The rapidly growing class of atomically thin  and  tunable van der Waals materials is  intensely investigated both in the context of  fundamental science and for new 
technologies. There is in this connection a widespread need for new ways to probe the electronic properties of these layered materials,  since their
two-dimensional (2D) character  make conventional probes less efficient. Here, we show how excitons trapped in a moir\'e lattice   can be used as an 
optical probe for spatially  resolved electron spin density correlations in such materials. The electrons in the material of interest virtually tunnel to the moir\'e lattice where 
they scatter on the excitons after which they tunnel back. This gives rise to an effective spin-dependent and spatially localised potential
felt by the electrons, which in turn leads to energy shifts that can be measured spectroscopically in the exciton spectrum. Using second order perturbation theory combined with a solution to 
the exciton-electron scattering problem, we show that the electrons mediate an interaction between two excitons resulting in an energy shift proportional to their two-point spin density-density correlation function evaluated at the exciton 
positions.  
We then discuss two specific applications of our setup. 
First, we show that quantum phase transitions between different in-plane anti-ferromagnetic orders in  a 2D lattice give rise to 
large and measurable shifts in the exciton spectrum in the critical regions.
Second, we analyse how different pairing symmetries of superconducting phases can be probed. 
This demonstrates that our scheme opens up new ways to probe  electron spin density correlations, which is a key property of  many quantum phases predicted to exist in the new 2D materials. 
 \end{abstract}

 \maketitle
\section{Introduction\label{Sec_intro}}
The increasingly sophisticated fabrication of atomically thin  transition metal dichalcogenides (TMDs) MX$_2$ (e.g.\ M$=$Mo or W and X$=$S, Se or Te) and
their  heterostructures have resulted in an increasing number of new and interesting 2D quantum materials~\cite{Wang2018}.
TMDs are direct band gap semiconductors that have band extrema at the so-called K and K' points in their hexagonal Brilllouin zone with a large spin-valley splitting, where 
each valley can be selectively excited using circularly polarized light~\cite{Mak2012,Zeng2012}. The reduced screening of the Coulomb interaction in these 2D materials combined  with their 
typically large electron and hole band masses lead to large interaction-to-kinetic energy ratios. One can moreover stack  TMDs on top 
of each other with a lattice mismatch or with a relative twist angle and in this way  
create a moir\'e superlattice potential with almost flat electronic bands in the reduced Brillouin zone. 
Many properties of these moir\'e systems can be tuned such as their carrier density and  ratio of interaction energy to hopping strength,  in order to realise new quantum many-body phases 
of interest for fundamental science and technological applications~\cite{Bistritzer2011,wu2018hubbard,pan2020band,zang2021hartree,Mak2022}. 
A major problem however is that transport spectroscopy, which is often used to probe strongly correlated systems, has proven difficult to implement for these  materials. Likewise, 
their layered structure makes them couple only weakly to $X$-ray and neutron spectroscopic probes. All this means that there is a widespread need for alternative ways to probe the  new 2D materials.  

One possibility is to use excitons as an optical probe for the electronic states of the 2D materials. The general idea is
that interactions with the electrons change the optical spectrum of the excitons 
 in ways that depend on the state of the electrons such as whether 
it is gapped or not. 
This  has been utilized to detect the formation of a Wigner crystal~\cite{salvador2022optical,Smoleski2021},   incompressible 
Mott states~\cite{Shimazaki2021}, spin ordering in a triangular moir\'e lattice~\cite{Ciorciaro2023}, and a range of  correlated phases~\cite{Jin2021,Xu2020,Miao2021,Zhou2021}. The  presence of quasiparticle peaks in the exciton spectrum was 
used to identify incompressible filled Landau levels in graphene~\cite{Cui2024}, quantum Hall states in TMDs~\cite{Smolenski2019}, and exciton insulators in a moir\'e lattice~\cite{Gu2022}. Theoretically, it has been analysed 
how different electronic phases show 
up  in the exciton spectrum~\cite{Mazza2022,julku2024exciton,Huang2023,Amelio2023,Sorout_2020}. 
For a recent review discussing  excitons as probes for 2D materials, see Ref.~\onlinecite{massignan2025polaronsatomicgasestwodimensional}.

Recently, resonant scattering between charge carriers in one layer and  excitons in another due to virtual tunneling has been 
observed experimentally~\cite{Experiments_tunable}, and this scattering process has been predicted to lead to magnetic 
polarons and string excitations for a mobile exciton~\cite{julku2024exciton}. Inspired by these results, 
we explore in this paper how density correlations in a 2D material can be probed in the optical spectrum of 
excitons trapped  in an adjacent moir\'e lattice. 
Electrons from the material of interest  tunnel virtually into the  moir\'e lattice, where they scatter on the excitons before tunneling back. Since an electron mainly scatters on an exciton with opposite spin, 
this gives rise to an effective, spin-dependent static potential for the electrons in the material. We  show that this can be used to measure the two-point spin density correlation
function of the electrons via the second order energy shift of the excitons. Using two excitons separated by a distance, the spatial dependence of the 
correlation function can be probed. 
Two specific examples of this are then discussed. We first demonstrate how 
the  quantum phase transitions between different in-plane anti-ferromagnetic orders in a 2D spin lattice give rise to a
large signal in the  exciton spectrum in the critical regions. Then we show how the spatial symmetry of the Cooper 
pairs in different superconducting phases can be probed using two excitons. 

\section{System\label{Sec_model}}
\begin{figure}[h]
\centering
\includegraphics[width=\columnwidth]{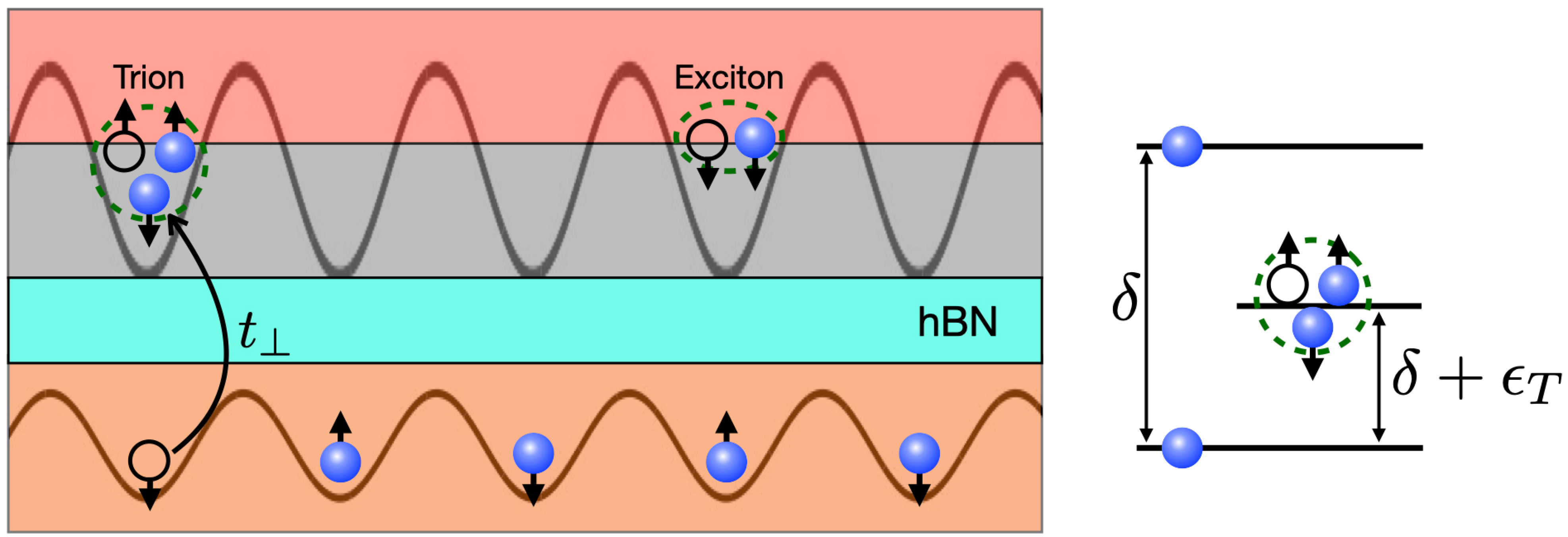}
\caption{Electrons (blue balls) in the lower material can tunnel to the upper material, where they scatter on 
excitons with opposite spin trapped by a deep moir\'e potential.  The exciton-electron interaction potential supports a  
bound state, i.e.\ a trion.
The energy levels of the electron in the lower material as well as of the electron and trion in the upper material  are shown on the right.}
\label{Fig-setup}
\end{figure}

We consider the setup illustrated in Fig.~\ref{Fig-setup}. An upper  system  contains 
 two excitons at positions ${\mathbf r}_1$ and ${\mathbf r}_2$, as can be achieved experimentally by trapping them in a deep moir\'e lattice~{\cite{seyler2019signatures, li2020dipolar, brotons2020spin}}. 
 The excitons have spins $\bar \sigma_1$ and $\bar \sigma_2$ meaning that they are formed by a spin $\bar \sigma=\uparrow,\downarrow$ 
 electron excited from a valance to a conduction band, which can be realised using the spin-valley locking of TMDs~\cite{Wang2018}. A lower system 
  contains electrons  whose properties we aim to probe with the excitons.  In the following, we refer to the upper and lower 2D systems as two layers even though they each typically consist of two or more layers forming 
  moir\'e lattices as in the examples we will consider below. We also refer to the charge carriers as electrons but they may be holes as well. 
  An insulating spacing material, such as hexagonal Boron Nitride (hBN), separates the two layers thereby prohibiting the formation 
  of a moir\'e system or other  hybridisation effects, while at the same time allowing for electron 
tunneling between the two layers. The Hamiltonian is 
\begin{align}
\hat H=&\sum_{\mathbf{k}\sigma}(\epsilon_{\mathbf k}+\delta)\hat a_{{\mathbf k}\sigma}^\dagger\hat a_{{\mathbf k}\sigma}
+\hat H_e
+t_\perp\sum_{\mathbf{k}\sigma}(\hat a_{{\mathbf k}\sigma}^\dagger\hat c_{{\mathbf k}\sigma}+\text{h.c.})
\nonumber\\
&+\frac1N\sum_{i=1}^2\sum_{{\mathbf k},{\mathbf k}'}e^{i({\mathbf k'}-{\mathbf k})\cdot {\mathbf r}_i}V_\text{ex}({\mathbf k}-{\mathbf k}')\hat a^\dagger_{\mathbf k\sigma_i}\hat a_{\mathbf k'\sigma_i},
\end{align}
where $\hat a_{\mathbf k\sigma}^\dagger/\hat c_{\mathbf k\sigma}^\dagger$ creates an electron in the upper/lower layer,  $\hat H_e$ is the Hamiltonian for the electrons in the lower layer, $\sigma$ is the opposite spin of $\bar \sigma$,
and $N$ is the number of lattice sites in the upper layer where the excitons are trapped. 
 The energy of an electron in the upper layer with  crystal momentum ${\mathbf k}$
 is $\epsilon_{\mathbf k}+\delta$ where $\delta$ is an energy off-set with respect to the 
 lower layer, which can be controlled by gating. 
 We have assumed that there is only a significant interaction $V_\text{ex}({\mathbf k})$ excitons with opposite spins in the upper layer, since the interaction between  electrons and excitons  with parallel 
 spins is strongly suppressed by the Pauli exclusion principle~\cite{massignan2025polaronsatomicgasestwodimensional}. 
 In particular, only the interaction between 
  electrons and excitons with opposite spins support a bound state, i.e.\ a trion, a property that will be used extensively below. 
   Finally, we have for simplicity assumed that the electron tunneling between the two layers with matrix element $t_\perp$ conserves crystal 
 momentum, which strictly speaking only holds when the (moir\'e) lattices in the upper and lower layers are identical. 
 In App.\ \ref{AppA} we show that when the lattice constants  $a_1$ and $a_2$ of the two layers differ,
the momentum can change by multiples of $\Delta\mathbf{k}\propto 1/a_1-1/a_2$ 
in the tunneling process. For a small lattice mismatch, these effects are however negligible so that 
the dominant tunneling process indeed conserves the in plane momentum. 
 
\section{Effective static potential}
\label{ScatSec}
\begin{figure}
\centering
\includegraphics[width=0.47\textwidth]{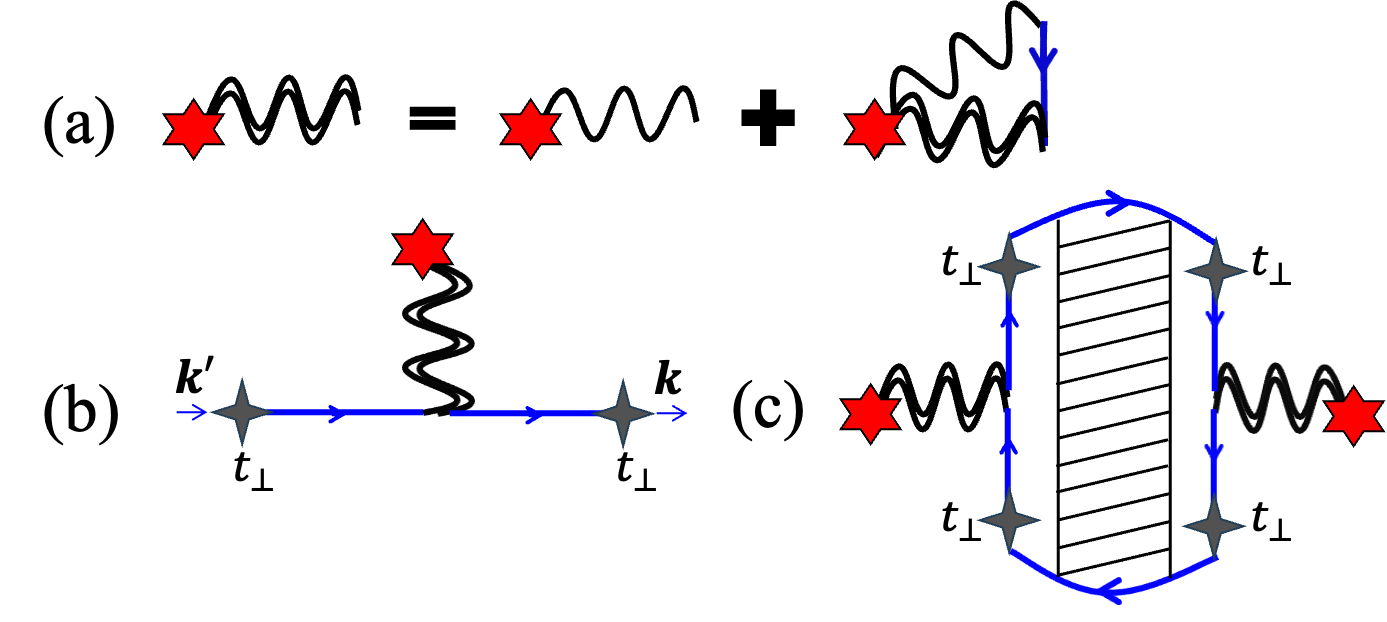}
\caption{(a) Diagrams for  electron-exciton scattering matrix in the upper layer. (b) Self-energy of the electrons in the lower layer due to tunneling to the upper layer, scattering on the exciton, and tunneling back. 
(c) The
interaction between two excitons mediated by electrons in the lower layer. 
Red stars (grey crosses) represent the static exciton (tunneling between two layers),  single (double) wavy lines are the bare electron-exciton interaction (electron-exciton scattering matrix), and the blue lines are the propagator for the electron. }
\label{Fig-diagram}
\end{figure}

We now show how the scattering of electron on an 
exciton in the upper layer gives rise to an effective static potential felt by the electrons in the lower layer.
For this purpose,  consider the scattering of an electron in the upper layer on a static exciton with opposite spin. The electron-exciton scattering matrix obeys the Lippmann–Schwinger equation [see Fig. \ref{Fig-diagram}(a)]
\begin{gather}
\mathcal{T}({\mathbf k},{\mathbf k}',\omega)=V_\text{ex}({\mathbf k}-{\mathbf k}')\nonumber\\
+\frac{1}{N}\sum_{\mathbf{k}''}V_\text{ex}({\mathbf k}-{\mathbf k}'')G_U(\mathbf{k}'',\omega)\mathcal{T}({\mathbf k}'',{\mathbf k}';\omega),
\label{Eq-T-up}
\end{gather}
where ${\mathbf k}/{\mathbf k}'$ is the outgoing/incoming momentum of the electron, $\omega$ is the scattering  energy, 
$G_U(\mathbf{k},\omega)=1/(\omega-\epsilon_{\mathbf k}-\delta)$ is the Green's function for an electron in the upper layer,
and  we have for notational simplicity taken the exciton to be located at the origin (${\mathbf r}_i=0$). 
We assume 
that the electron can bind the  exciton to form a moir\'e trapped trion with energy $\epsilon_T<0$  relative to the exciton energy
as  observed experimentally~\cite{brotons2021moire, baek2021optical, liu2021signatures}. It follows that the scattering matrix diverges at the energy 
$\omega=\delta+\epsilon_T$. For a short range electron-exciton interaction this can  be used to 
replace $V_\text{ex}({\mathbf k})$ with the trion energy in the Lippmann-Schwinger equation~\cite{massignan2025polaronsatomicgasestwodimensional}. We can to a good approximation  ignore 
the frequency dependence of the scattering matrix for $\omega\ll\delta+\epsilon_T$, which together with 
$G_U(\mathbf{k},\omega)\simeq-1/\delta$ means that the solution to the Lippmann-Schwinger equation is   
\begin{equation}
\mathcal{T}({\mathbf k},{\mathbf k}',\omega)\simeq\frac{\epsilon_T\delta}{\delta+\epsilon_T}.
\label{Eq-T-approx}
\end{equation}
Details of this derivation are given in App.~\ref{AppB}. When the exciton is located at position ${\mathbf r}_i$ away from the origin, the 
scattering matrix in Eq.~\eqref{Eq-T-approx} is simply multiplied by a factor $e^{i({\mathbf k}'-{\mathbf k})\cdot {\mathbf r}_i}$.
Note that Eq.~\eqref{Eq-T-approx} breaks down when the energy $\omega$ of the electron is close to that of the trion 
$\epsilon_T+\delta$, which  gives rise to resonant scattering with a strong energy dependence,  see App.~\ref{AppB}.

We can now calculate the static  potential experienced by spin $\sigma$ electrons in the lower layer due to an exciton 
in the upper layer with opposite spin $\bar\sigma$ located at  ${{\mathbf r}_i}$. 
The tunneling  between the two layers gives rise to a  self-energy term for $\sigma$ electrons in the lower layer  of the form 
\begin{eqnarray}
\Sigma({\mathbf k,\mathbf k',\omega})&=&
t^2_{\perp}G_U(\mathbf{k},\omega)\mathcal{T}({\mathbf k,\mathbf k',\omega})G_U(\mathbf{k}',\omega)\nonumber\\
&&+t_{\perp}^2G_U(\mathbf{k},\omega)\delta_{\mathbf{k},\mathbf{k}'}.
\label{Eq-V-wk}
\end{eqnarray}
Physically, Eq.~\eqref{Eq-V-wk} describes a spin $\sigma$ electron with crystal momentum $\mathbf{k}'$ in the lower layer tunneling to the upper layer where it scatters on the 
$\bar\sigma$  exciton to momentum $\mathbf{k}$ after which it tunnels down to the lower layer again as illustrated diagrammatically 
in Fig. \ref{Fig-diagram}(b). 
The second term in Eq.~\eqref{Eq-V-wk}  corresponds to free electron propagation in the upper layer without scattering, which gives a constant energy shift that is omitted in the following. 
 
Assuming again that the typical energy $\omega$ of the electrons in the lower layer is far detuned from the scattering resonance, i.e.,    $\omega\ll\delta+\epsilon_T$, 
the energy/momentum dependence of scattering matrix can be ignored so that  it is given by Eq.~\eqref{Eq-T-approx}. Using this together with  
$G_U(\mathbf{k},\omega)\simeq-1/\delta$, the self-energy in Eq.~\eqref{Eq-V-wk} reduces to that coming from 
a static scattering potential given by 
\begin{align}
\hat V_\text{eff}(\mathbf r_i)&= \frac{t^2_{\perp}}{\delta^2}\frac{\epsilon_T\delta}{\delta+\epsilon_T}\frac 1N\sum_{\mathbf k,\mathbf k'}e^{i(\mathbf k'-\mathbf k)\cdot \mathbf r_i}
\hat c^\dagger_{\mathbf k\sigma}\hat c_{\mathbf k'\sigma}\nonumber\\
&=\frac{t^2_{\perp}}{\delta}\frac{\epsilon_T}{\delta+\epsilon_T}\hat c^\dagger_{i\sigma}\hat c_{i\sigma}=\kappa \hat n_{\sigma}({\mathbf r}_i).
\label{Veff}
\end{align}  
Here, $\hat c^\dagger_{i\sigma}=N^{-1/2}\sum_{\mathbf k}e^{-i{\mathbf k}\cdot{\mathbf r}_i}\hat c^\dagger_{\mathbf k\sigma}$ 
creates a spin $\sigma$ electron at position ${\mathbf r}_i$, and $\hat n_{\sigma}(\mathbf r_i)=\hat c^\dagger_{i\sigma}\hat c_{i\sigma}$
is the density of $\sigma$ electrons at position ${\mathbf r}_i$. Consistent with assuming momentum conservation in the tunneling between the 
layers, we have taken the lattices in two layers to be identical when performing the Fourier transformation. Equation \eqref{Veff} shows that a spin $\bar\sigma$  exciton at position  ${\mathbf r}_i$ in the upper layer gives rise to a  local 
 potential experienced by the spin $\sigma$  electrons in the lower layer with a strength 
$\kappa=t^2_{\perp}\epsilon_T/[\delta(\delta+\epsilon_T)]$. We emphasize that the same scattering process as that considered here involving virtual tunneling of charge carriers in one layer to another layer containing an exciton has recently been experimentally realised~\cite{Experiments_tunable,kuhlenkamp2022tunable}. In this case, resonant 
hole-exciton scattering was achieved by tuning the energy offset between the layers to cancel the trion energy, which corresponds to setting
$\epsilon_T+\delta=0$ in the present case.

\section{Probing spin density correlations}
Having derived the effective static potential felt by  electrons in the lower layer due to trapped excitons in the upper layer, we will in this 
 section show how this can be used to realise an optical probe of spin density correlations of the electrons.

Consider first a single  exciton at position ${\mathbf r}_1$ in the upper layer with spin $\bar \sigma$. 
This leads to a local  
potential given by Eq.~\eqref{Veff} for the spin $\sigma$ electrons below. Assuming that this potential is weak, 
 the resulting energy shift   can be calculated 
using perturbation theory. The first order  energy shift at zero temperature is  given by the mean-field expression 
\begin{equation}
E_1=\langle \psi_0|\hat{V}_{\text{eff}}|\psi_0\rangle=\kappa n_\sigma({\mathbf r}_1),
\label{Eq-E1-1X}\end{equation}
where $|\psi_0\rangle$ is the many-body ground state of the electrons in the lower layer with spin density $ n_\sigma({\mathbf r}_1)=\langle\psi_0|\hat n_\sigma({\mathbf r}_1)|\psi_0 \rangle$. This shows that the energy shift 
of the exciton   can be used to measure spin resolved densities in the lower layer. For instance, 
an out-of-plane ferromagnetic order can be detected spectroscopically as an exciton energy shift depending on its spin, and an out-of-plane anti-ferromagnetic 
order will show up as an  energy shift oscillating as a function of the exciton position. 
For a non-zero temperature, we can simply replace $\langle \psi_0|\ldots|\psi_0\rangle$ in Eq.~\eqref{Eq-E1-1X} by a thermal average
$\langle\ldots\rangle$.

Probing  the spin density \emph{correlations} of electrons, which often provide smoking gun signals of their quantum state,  remains a major challenge for 2D materials. 
We now show how this can be achieved by measuring the second order energy shift of the excitons. Consider
 two excitons at positions ${\mathbf r}_i$ with spins $\bar \sigma_i$ ($i=1,2$)
 giving rise to a first order energy shift  $E_1=\kappa n_{\sigma_1}({\mathbf r}_1)+\kappa n_{\sigma_2}({\mathbf r}_2)$. 
The second order shift is
\begin{align}
E_2=&\sum_{n>0}\frac{|\langle \psi_n|\hat{V}_{\text {eff}}({\mathbf r_1})+\hat{V}_{\text {eff}}({\mathbf r_2})|\psi_0\rangle|^2}{E_{0}-E_n}\nonumber\\
= &-\frac{\kappa^2}2\left[\langle\langle \hat n_{\sigma_1}({\mathbf r}_1),\hat n_{\sigma_1}({\mathbf r}_1)\rangle\rangle+\langle\langle \hat n_{\sigma_2}({\mathbf r}_2),\hat n_{\sigma_2}({\mathbf r}_2)\rangle\rangle\right]\nonumber\\
 &-\kappa^2\langle\langle \hat n_{\sigma_1}({\mathbf r}_1),\hat n_{\sigma_2}({\mathbf r}_2)\rangle\rangle,
\label{SecondOrder}
\end{align}
where $|\psi_n\rangle$ is an eigenstate of the electron system with energy $E_n$ and $E_0$ is the ground state energy. Here, 
$\langle\langle\hat A,\hat B\rangle\rangle$ 
denotes the zero frequency component of the retarded  correlation function
\begin{equation}
\langle\langle\hat A,\hat B\rangle\rangle(t)=-i\theta(t)\langle[\hat A(t),\hat B(0)]\rangle
\end{equation}
where $\hat O(t)=e^{i\hat H_et}\hat Oe^{-i\hat H_et}$ is a time-dependent operator in the Heisenberg representation with respect to the 
many-body Hamiltonian of the electrons in the lower layer, see App.~\ref{AppC}. 

Equation \eqref{SecondOrder} shows that the second order energy shift of the two excitons is proportional to the sum of spin resolved 
density-density correlations of the electrons. It follows that 
one can   probe the spatial and spin dependence of the electron density correlations  in a material of interest
using optical spectroscopy of excitons. 
The appearance of $\langle\langle \hat n_{\sigma_1}({\mathbf r}_1),\hat n_{\sigma_2}({\mathbf r}_2)\rangle\rangle$
 can be interpreted as the interaction between the two excitons in the top layer mediated by the electrons in the 
lower layer as illustrated diagrammatically in Fig.\ \ref{Fig-diagram}(c). A similar approach has been used to derive  the 
interaction between  static ions and neutral particles in ultracold atomic gases~\cite{Shanshan2022,Paredes2024}. 
Equation \eqref{SecondOrder} can straightforwardly be 
generalised to the case of an arbitrary number of excitons at positions ${\mathbf r}_i$ in the lattice.

\section{Probing magnetic phases}\label{MagneticSection}
We now consider a  specific use case of our setup: 
 probing spin correlations in different magnetic phases of electron 
lattice systems. Consider a half filled lattice with strong on-site repulsion between the electrons so that their  
low energy spin degrees of freedom are  described by a Heisenberg model~\cite{auerbach2012interacting}. 
Since there is one electron per site, we can write $\hat{n}_{\sigma}({\mathbf r}_i)=1/2+\sigma \hat S_z({\mathbf r}_i)$ where 
$\hat{\mathbf{S}}({\mathbf r}_i)=[\hat S_x({\mathbf r}_i),\hat S_y({\mathbf r}_i),\hat S_z({\mathbf r}_i)]$ is the vector operator for the 
spin 
at position  
${\mathbf r}_i$ and $\sigma=\pm$ for spin $\uparrow/\downarrow$ in this section.
It immidiately follows that 
\begin{equation}
\langle\langle \hat n_{\sigma_i}({\mathbf r}_i),\hat n_{\sigma_j}({\mathbf r}_i)\rangle\rangle=\sigma_i\sigma_j\langle\langle \hat S_z({\mathbf r}_i),\hat S_z({\mathbf r}_j)\rangle\rangle,
\label{SpinDensityCorr}
\end{equation}
 which together with  Eq.~\eqref{SecondOrder} demonstrates that  
 our scheme is well-suited for probing the spin correlations. 

\begin{figure}
\centering
\includegraphics[width=0.47\textwidth]{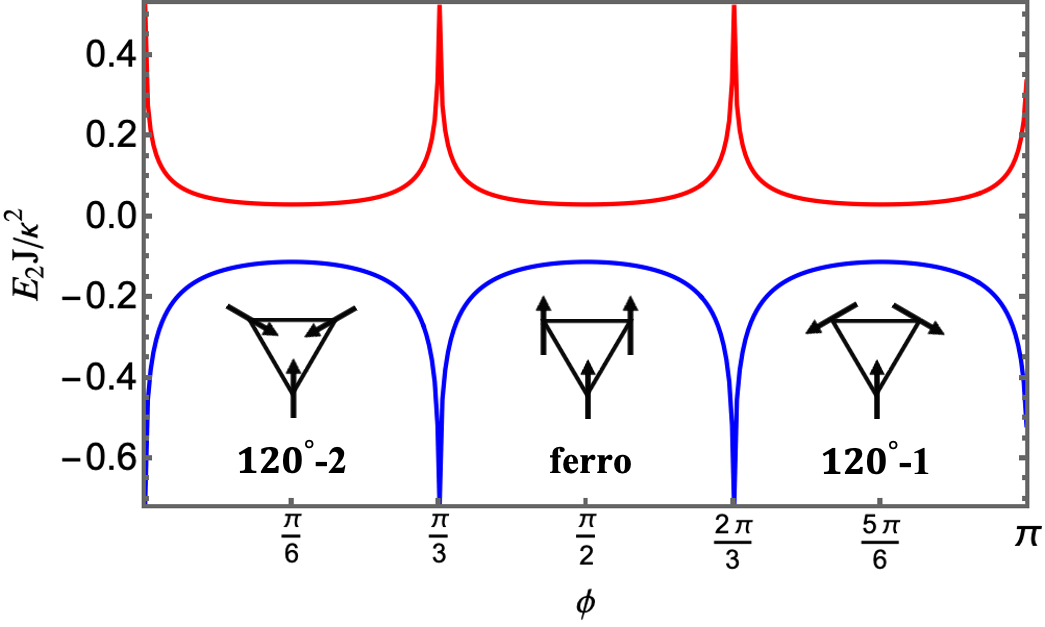}
\caption{The second order energy shift of the spin model given by Eq.\ \eqref{Eq-H} due to the excitons in the upper layer. The blue line represents the result for one exciton 
and the red line shows  the interaction term between  two nearest neighbour excitons with the same spin. }
\label{Fig-E20-E2R}
\end{figure}

Take as a concrete example a lower system consisting of  a twisted bilayer WSe$_2$, 
which in the strong repulsion limit and at half filling is described by the 
spin model~\cite{pan2020band, zang2021hartree}
\begin{gather}
\hat{H}_S=J\sum_{\langle i,j\rangle}\big\{\hat S_z({\mathbf r}_i)\hat S_z({\mathbf r}_j)+\cos2\phi_{ij}[\hat S_x({\mathbf r}_i)\hat S_x({\mathbf r}_j)\nonumber\\
+\hat S_y({\mathbf r}_i)\hat S_y({\mathbf r}_j)]
+\sin2\phi_{ij}[\hat{\mathbf{S}}({\mathbf r}_i)\times \hat{\mathbf{S}}({\mathbf r}_j)]\cdot \hat{z}\big\}
\label{Eq-H}\end{gather}
on a  triangular lattice. Here,   $J>0$ is the superexchange coupling between neighbouring spins and  
the phase $\phi_{ij}$ is  symmetric under $2\pi/3$ rotations with $\phi_{ij}=-\phi_{ji}$. 
 For convenience, we define $\phi_{ij}=\phi$ for the phase 
 between a spin at site $i$ and the  nearest neighbour spin $j$ in the horizonthal ($x$) direction.
 The ground state of $\hat{H}_S$ is  a $120^\circ-2$ in-plane antiferromagnet, an in-plane ferromagnet, and a $120^\circ-1$ in-plane antiferromagnet  when $\phi\in [0, \pi/3]$, 
$[\pi/3, 2 \pi/3]$, and $[2\pi/3, \pi]$ respectively~\cite{zang2021hartree}, as illustrated in Fig.~\ref{Fig-E20-E2R}. Since the magnetic order  is in-plane,  
the first order energy shift Eq.~\eqref{Eq-E1-1X} will be the same for all three phases. In order to distinguish them, 
 we therefore focus on the second order energy shift given by 
Eq.~\eqref{SecondOrder}. 

 The spin-spin correlation function  is within linear spin-wave theory given by 
 \begin{align}
\langle\langle \hat S_z(\mathbf{r}_i),\hat S_z(\mathbf{r}_j)\rangle\rangle=\frac{1}{NJ}\sum_\mathbf{k}
\frac{e^{i\mathbf{k}\cdot(\mathbf{r}_i-{\mathbf r}_j)}}{A_\mathbf{k}+B_\mathbf{k}}
\label{SpinSpin}
\end{align}
for the in-plane ferromagnetic phase. Here
\begin{align}
A_\mathbf{k}&=-6\cos 2\phi+\left(\cos 2\phi+1\right)\gamma_\mathbf{k},\\
B_\mathbf{k}&=-\left(\cos 2\phi-1\right) \gamma_\mathbf{k},\\
\gamma_\mathbf{k}&=\cos k_x+2\cos(k_x/2)\cos(\sqrt{3}k_y/2),\label{Gamma}
\end{align}
and we have taken the lattice constant to be unity. 
The spin-spin correlation functions for the $120^\circ-1$ and $120^\circ-2$ antiferromagnetic phases 
are same as above except for a shift of the angle $\phi$ by $\pm \pi/3$. Details of deriving Eqs.~\eqref{SpinSpin}-\eqref{Gamma} are
given in  App.~\ref{AppD}.

The blue line in Fig.~\ref{Fig-E20-E2R} shows the second order energy shift of a single exciton as a function of the angle $\phi$ given by 
Eq.~\eqref{SecondOrder} and Eq.~\eqref{SpinDensityCorr} without the 
terms involving $n_{\sigma_2}(\mathbf{r}_2)$, where we have evaluated Eq.~\eqref{SpinSpin} numerically.
Figure \ref{Fig-E20-E2R} shows  that interactions of the exciton with the electrons below give rise to a negative second order energy shift of the order  $\kappa^2/J$. We also see that 
the energy shift increases as the transitions between the different anti-ferromagnetic phases are approached, and that it 
diverges at the  critical points. These divergencies arise because the different phases differ by spin rotations of the spins in the $xy$-plane, which is  generated by $\hat S_z$, or equivalently because $\hat S_z$ 
excites spin waves of the magnet, which become soft 
at the phase transition points.  Of course, second order perturbation theory 
is inaccurate for such large energy shifts and a non-perturbative approach is therefore needed in the critical regions  to get quantitatively accurate results. 
The result that magnetic phase transitions give rise to a large signal in the exciton spectrum is 
nevertheless robust, since the excitons probe spin-spin correlation functions that typically diverge at the critical points.

\begin{figure}
\centering
\includegraphics[width=0.47\textwidth]{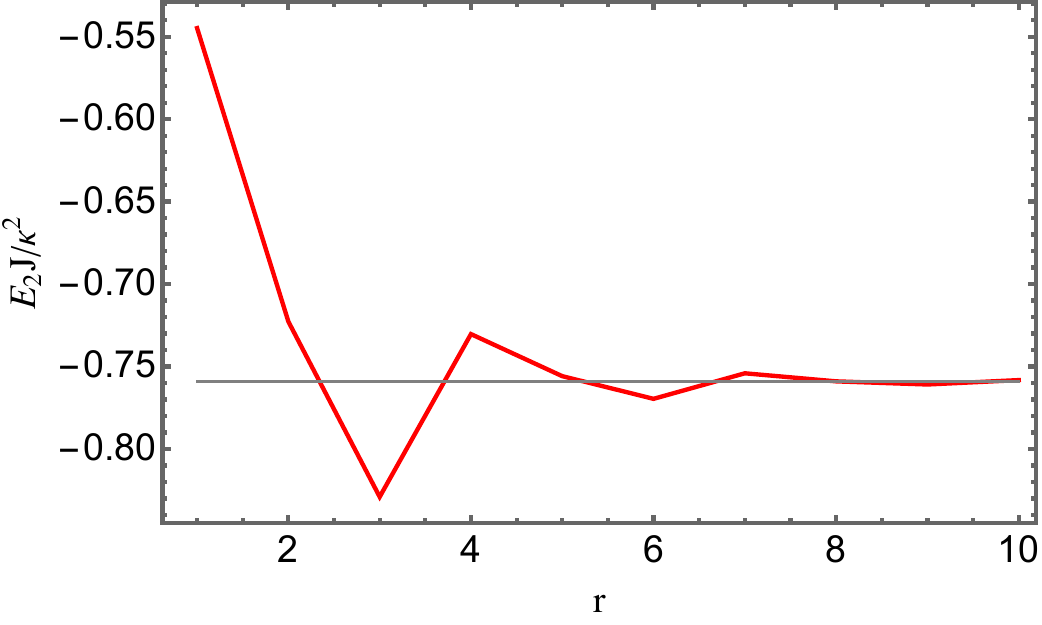}
\caption{The  second order energy shift for two  excitons with parallel spins in the upper layer as a function of their separation 
$r$ in units of the lattice constant with $\phi=0.34\pi$.
}
\label{Fig-E2R}
\end{figure}

Consider now two excitons with parallel spins located at nearest neighbour lattice sites $\mathbf r_1$ and $\mathbf r_2$. 
The red line in Fig.~\ref{Fig-E20-E2R} shows the ${\mathbf r}=\mathbf r_1-\mathbf r_2$ dependent term of the second order 
energy shift of two excitons  given by the last term in Eq.~\eqref{SecondOrder} proportional to 
$\langle\langle \hat S_z(\mathbf{r}_1),\hat S_z(\mathbf{r}_2)\rangle\rangle$.
  This shows that the interaction between  
two nearest neighbour excitons with parallel spins mediated by the spins is repulsive and that it also diverges at the phase transitions. 
To examine the spatial dependence further, we plot in Fig.~\ref{Fig-E2R} the energy shift of the two excitons as a function of their separation $r$. 
We have taken $\phi=0.34\pi$ so that the 
system is in the ferromagnetic phase close to the transition point to the anti-ferromagnetic $120^\circ-2$ phase, see Fig.~\ref{Fig-E20-E2R}. 
The energy shift due to the 
interaction mediated by spin waves oscillates as a function of the distance between the two excitons akin to the  Friedel oscillations of the Ruderman-Kittel-Kasuya-Yosida (RKKY) 
 interaction~\cite{ruderman1954indirect, kasuya1956theory, yosida1957magnetic}. 
 For large distance $r\gg 1$ between the two excitons, the energy shift reduces to that of two isolated excitons (grey line) as expected. By measuring these oscillations, one  can 
 therefore probe the  spatial dependence of the spin-spin correlations of the electrons  or equivalently the spin-wave mediated interaction. 
  The energy shift has a similar behaviour for other values of $\phi$  with the overall magnitude increasing  the closer the system is to a critical point 
  separating two magnetic phases. 

Taking  experimentally realistic values for the spin coupling 
$J=0.05$meV~\cite{tang2020simulation} and the trion energy $\epsilon_T=-5$meV\cite{liu2021signatures, baek2021optical, brotons2021moire}, assuming the energy off-set 
between the two layers is $\delta=5.5$meV, and using $t_\perp=1$meV for the interlayer tunneling matrix element, 
we get $\kappa^2/J\simeq66$meV. This indicates that the energy shifts reported in 
Figs.~\ref{Fig-E20-E2R}-\ref{Fig-E2R} should be  experimentally observable. 
   From this we conclude that our setup can be applied to probe spin correlations  of strongly 
 correlated electrons in the Mott limit including their spatial dependence. 
 This is useful since such correlations are a key property characterizing many strongly correlated phases also 
  without magnetic order such as  spin liquids~\cite{zhou2017quantum}. Also, 
  since  magnetic phase transitions in general give rise to diverging spin-spin correlation functions they can be detected optically 
by a corresponding large energy shift in the exciton spectrum. 


\section{Probing  superconducting phases}
\label{section:Supercond}
As another application of our setup, we now explore the probing of different superconducting phases. Consider
a simple BCS model for a nearest-neighbor tight-binding model on a triangular lattice with the Hamiltonian 
\begin{equation}
\hat H_\text{BCS}=\sum_{\mathbf k\sigma}\xi_{\mathbf k}\hat c_{\mathbf k\sigma}^\dagger\hat c_{\mathbf k\sigma}
+ \sum_{\mathbf k}(\Delta_{\mathbf k}\hat c_{\mathbf k\uparrow}^\dagger\hat c_{-\mathbf k\downarrow}^\dagger+\text{h.c.}). 
\label{Hbcs}
\end{equation}
Here, $\xi_{\mathbf{k}}=-2t[\cos k_{x}+2\cos (k_{x}/2)\cos(\sqrt{3}k_{y}/2)]-\mu$ is the tight binding kinetic energy for a triangular lattice with $t$ the nearest neighbor hopping matrix element, $\mu$  the chemical potential of the electrons and $\Delta_{\mathbf k}$  the pairing strength between time-reversed electrons. Pairing will mainly affect the density correlations between opposite spin electrons, which show up in the density-density correlation functions entering  Eq.~\eqref{SecondOrder}.
We find from a straightforward diagonalization of Eq.~\eqref{Hbcs}(see App.~\ref{AppE}) 
\begin{align}
\langle\langle \hat n_{\downarrow}(\mathbf{r}_1),\hat n_{\uparrow}(\mathbf{r}_2)\rangle\rangle
   =\frac{1}{2N^{2}}\sum_{\mathbf{k,k'}}\frac{\Delta_{-\mathbf{k}}\Delta^{*}_{\mathbf{k'}}}{E_{\mathbf{k}}+E_{\mathbf{k'}}}\frac{e^{-i(\mathbf{k+k'})\cdot \mathbf{r}}}{E_{\mathbf{k}} E_{\mathbf{k'}}}
     \label{ec:antiparallel_supercond}
 \end{align}
 where ${\mathbf r}=\mathbf{r}_1-\mathbf{r}_2$, 
 $E_{\mathbf{k}}=\sqrt{\xi_k^2+|\Delta_{\mathbf k}|^2}$.
 Equation \eqref{ec:antiparallel_supercond}  shows as expected  that Cooper pairing gives rise to density correlations between anti-parallel spins,  
 which in turn can be  probed via the resulting second order energy shift of two excitons with anti-parallel spins. From Eqs.~\eqref{SecondOrder} and \eqref{ec:antiparallel_supercond} it follows that the energy shift scales  as 
 $\kappa^2\Delta_{{\text{BCS}}}^2/t^3$.

 To demonstrate this, we consider two excitions with  antiparallel spins separated by a distance $r$ interacting with superconducting electrons in a triangular lattice. 
 We choose  three different kinds of pairings with the   symmetry of the  triangular lattice, 
 namely~\cite{Possible_symmetries}
\begin{equation*}
\frac{\Delta_{\mathbf k}}{\Delta_0}=
\begin{cases}
    1&s\text{-wave}\\
    2i[\sin k_{x}+\cos(\sqrt{3}k_{y}/2)\sin(k_{x}/2)]/\sqrt3&\text{chiral}\\
    \quad+2\sin(\sqrt{3}k_{y}/2)\cos(k_{x}/2) &p\text{-wave}\\
    2[\cos k_{x}-\cos(k_{x}/2)\cos(\sqrt{3}k_{y}/2)]/\sqrt{3}&\text{chiral} \\
    \quad-2i\sin(\sqrt{3}k_{y}/2)\sin(k_{x}/2)&d\text{-wave}
\end{cases}
\end{equation*}
where $\Delta_0$ is the overall magnitude of the pairing. 

\begin{figure}
    \centering
    \includegraphics[width=1.0\linewidth]{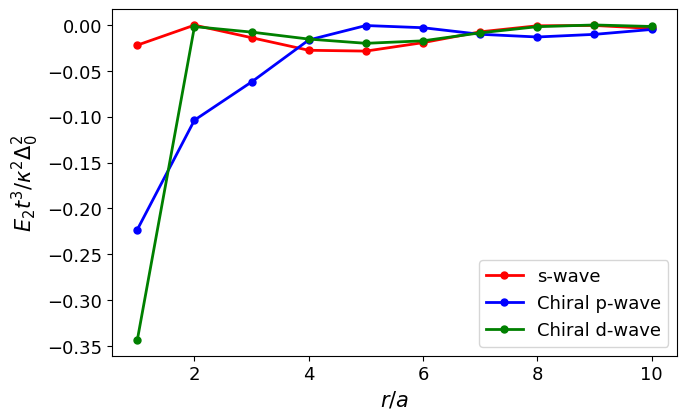}
    \caption{The energy shift of two excitons with anti-parallel spins  interacting with superconducting samples having three different kinds of Cooper pairing as a function of their separation. We have subtracted the energy  shift when the 
    excitons are infinitely apart. 
    }
    \label{fig:antiparallel-spins}
\end{figure}

Figure \ref{fig:antiparallel-spins} shows the $r$-dependent part of the second order energy shift of two excitons with anti-parallel spins
due to interactions with a superconductor with the pairing symmetries given above.  We have for the numerical calculations taken $\Delta_{0}/t=0.1$ and used a  chemical potential $\mu/t=0.83$ corresponding to 
an approximately half-filled ($49.93\%$) lattice, where superconductivity has been observed both in twisted bilayer graphene and in twisted bilayer WSe$_{2}$~\cite{tungsten_supercond,graphene_supercond}. The energy shift is generally negative reflecting that   Cooper pairing leads to positive density correlations between opposite spins. The spatial 
dependence of the energy shift is moreover determined by the gap function $\Delta_{\mathbf k}$   as 
quantified by Eq.~\eqref{ec:antiparallel_supercond}.

\begin{figure}
    \centering
    \includegraphics[width=0.8\linewidth]{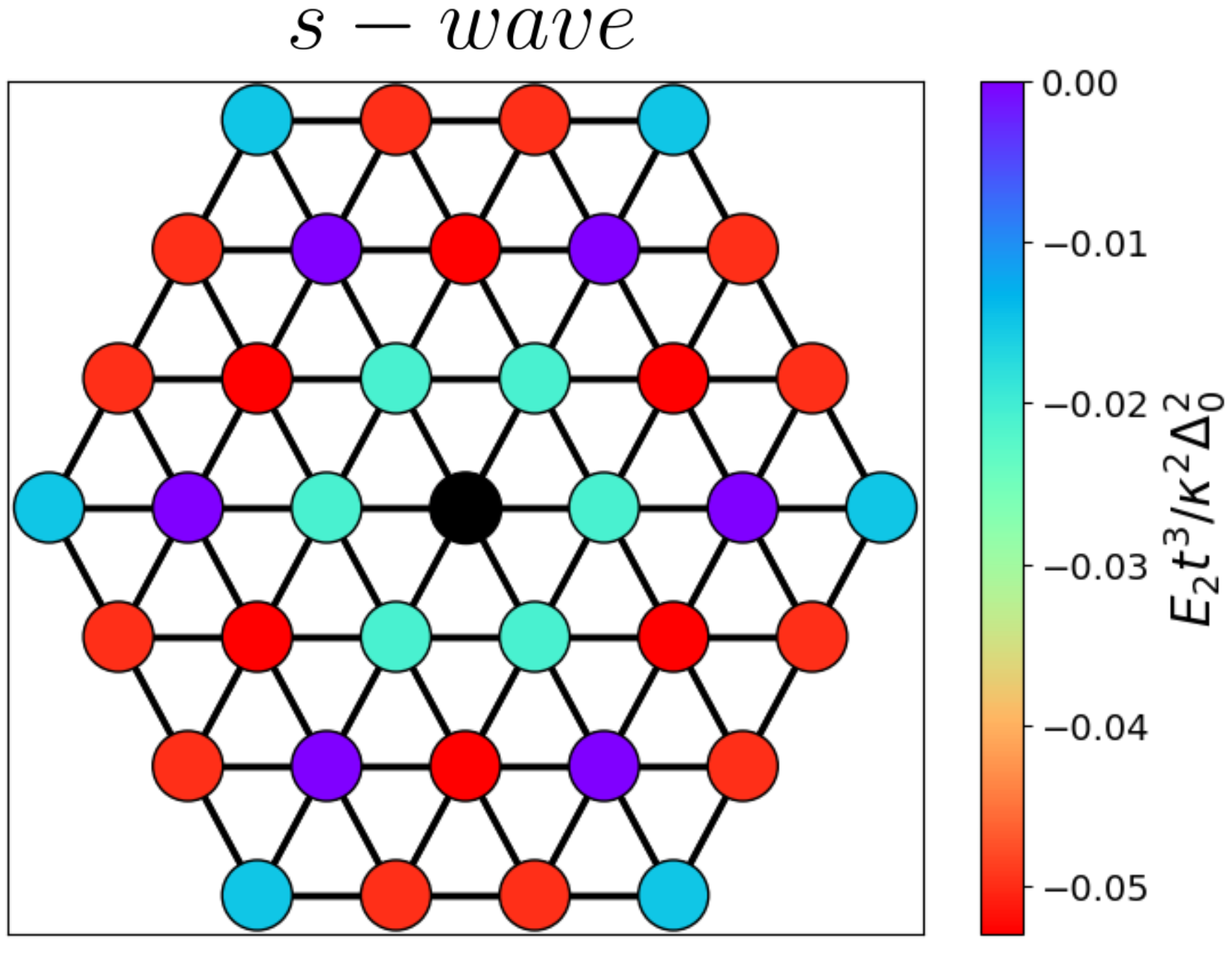}
    \includegraphics[width=0.8\linewidth]{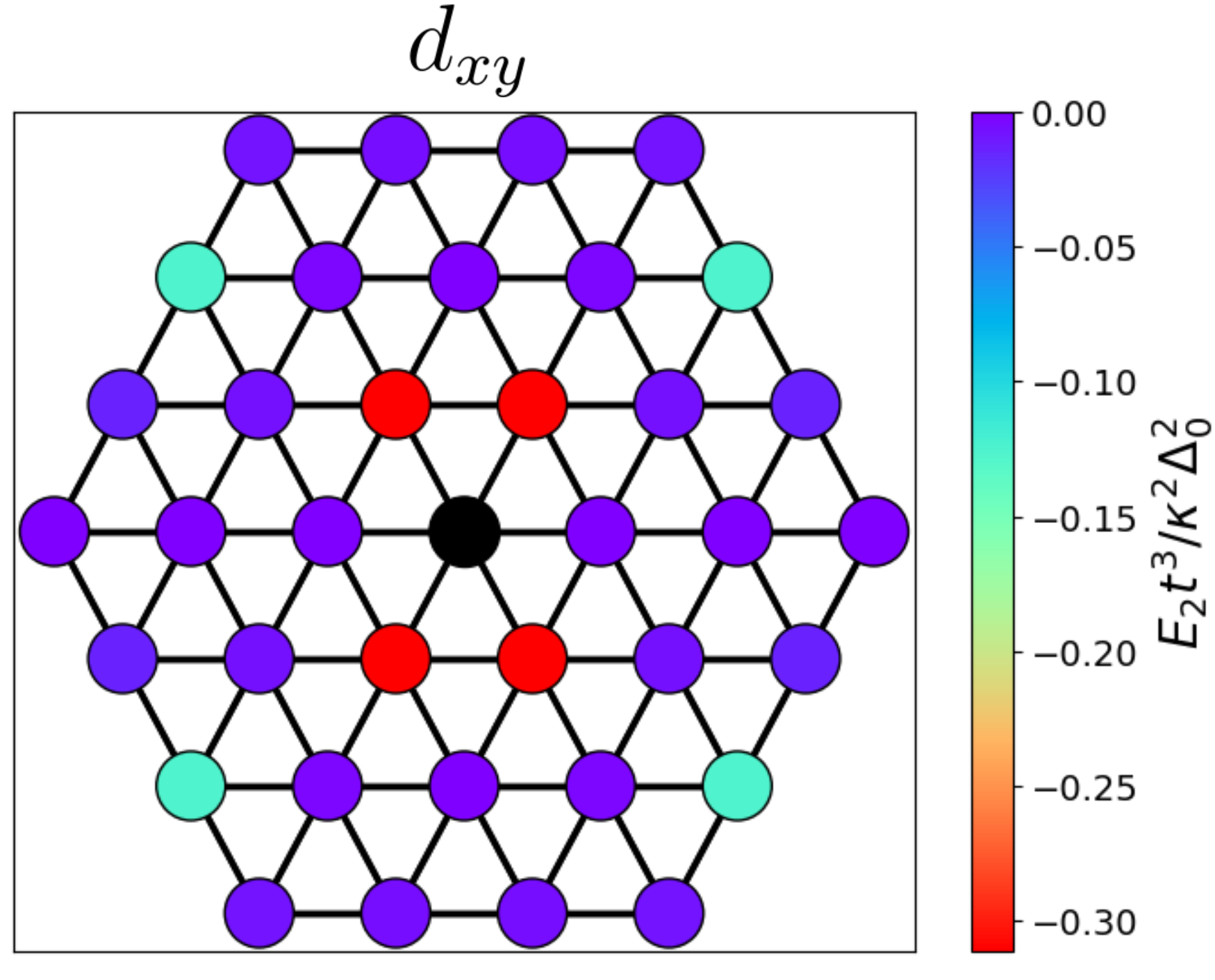}
    \caption{The  second order energy shift of two excitons with anti-parallel spins for a $s$-wave 
    superconductor (top) and a $d_{xy}$ superconductor (bottom). One exciton is at the origin (black dot) and the color codes give the 
    energy shift when the other exciton is at a given lattice site. The energy shift when the two excitons are infinitely apart is subtracted.}
    \label{fig:Symmetrybreak}
\end{figure}

To further illustrate how the spatial properties of the Cooper pairs can be probed, 
we plot in Fig.~\ref{fig:Symmetrybreak} the $\mathbf r$ dependent part of the second order energy shift 
in the full lattice for $s$-wave pairing and $d_{xy}$-wave pairing with 
$\Delta_{\mathbf{k}}=2\Delta_{0}\sin(\sqrt{3}k_{y}/2)\sin(k_{x}/2)$. We have chosen the latter in order to showcase results for 
Cooper pairs breaking the underlying lattice symmetry.
The plots  clearly illustrate how the spatial structure of the Cooper pairs  shows up in the dependence of 
the energy shift of the two excitons on their spatial separation. 

To estimate the magnitude of the energy shifts in Figs.~\ref{fig:antiparallel-spins}-\ref{fig:Symmetrybreak}, we assume that 
 the underlying band structure of the superconductor has a relatively small bandwidth of $5$meV
 as in twisted bilayer graphene \cite{graphene_supercond}. Since the bandwidth in  a triangular tight-binding model is $9t$, 
 this corresponds to $t\simeq 0.56$meV. Setting $\Delta_0/t=0.1$ corresponding to a pairing strength 
 90 times smaller than the band width, and using the same values for the  interlayer hopping  $t_{\perp}= 1$meV and the trion energy 
 $\epsilon_{T}= -5$meV as in Sec.~\ref{MagneticSection}, we obtain 
$\kappa^2\Delta_0^2/t^3\simeq 0.41$meV for $\delta=5.2$meV. This makes the energy shifts reported 
in Figs.~\ref{fig:antiparallel-spins}-\ref{fig:Symmetrybreak} at the limit of what can be observed. The magnitude of the 
energy shifts can of course  be increased by tuning the energy off-set $\delta$ so that the exciton-electron 
scattering is closer to resonance. A reliable calculation of the energy shift in this regime however requires an extension of our theory to include
 the frequency dependence of the scattering omitted in Eq.~\eqref{Eq-T-approx}. This is left for future work. 
 These results indicate that  our setup can be used to probe the properties of superconducting phases including the symmetry of the Cooper pairs. We have here used a minimal  BCS Hamiltonian to demonstrate proof-of-concept of our scheme, and one should  apply more  realistic but also more complicated Hamiltonians to explore this in more detail for specific superconductors.  Such models should  in particular
include non-superfluid correlations between anti-parallel spins, which in general are present also in the absence of pairing.

\section{Discussion and outlook}
We have demonstrated that excitons trapped in a moir\'e lattice can serve as an optical probe of electron spin density correlations
 in an adjacent 
2D material. Virtual electron tunneling between the material and the moir\'e lattice combined with  electron-exciton scattering 
gives rise to a spin-dependent static potential felt by the electrons in the 2D material. To second order, this in turn affects the energy
by an amount proportional to the spin density-density correlation  function of the electrons. By measuring the energy shift as a 
function of the separation between two excitons, one can therefore  extract the spatial dependence of this correlation function, which  
can be interpreted as an 
interaction between the two excitons mediated by the electrons in the material. 
We then discussed two concrete applications of this 
setup: Detecting quantum phase transitions between different in-plane anti-ferromagnetic phases and measuring the spatial symmetry of Cooper pairs in 
different superconductors. 

Our scheme can be implemented experimentally for instance by creating excitons in a moir\'e lattice with a certain 
concentration. This will give rise to a range of exciton spectral lines coming from excitons with different numbers of other excitons in their surroundings.
These specral lines may then be identified with e.g.\ a statistical analysis combined with theoretical modeling. Since the 
mediated interaction between two excitons generally decreases with their distance, it will likely in most cases be sufficient to consider exciton pairs separated 
by maximally   a few lattice sites.

These results demonstrate that our scheme can probe spin density correlations between electrons with spatial 
resolution. This is useful since such correlations  are a key  property of strongly interacting electronic phases that often
are difficult to measure by other means in 2D materials. One can tune several parameters 
in order to increase the sensitivity  including the  interlayer tunneling 
strength $t_\perp$ and the layer detuning $\delta$. As discussed above, one should however 
note that tuning $\delta+\epsilon_T$ close to zero to increase the 
signal will result in resonant exciton-electron scattering with a strong frequency dependence. Including this frequency dependence to calculate 
the sensitivity of our setup for resonant interactions is an interesting future problem. Another fascinating research direction
is to explore the use of more than two excitons 
to probe higher order correlation functions. Such high order correlation functions  provide deep insights into strongly correlated phases, and they are generally very hard to access by other means. 

\section{Acknowledgments}
This work has been supported by the Danish National
Research Foundation through the Center of Excellence
“CCQ” (Grant Agreement No. DNRF156).
Z.\ W.\ acknowledges support from National Key R$\&$D Program of China (Grant No. 2022YFA1404103), Natural Science Foundation of China (Grant No.~12474264) and Guangdong Provincial Quantum Science Strategic Initiative (Grant No.~GDZX2404007). S.D. is also supported by the Fundamental Research Funds for the Central Universities and the National Natural Science Foundation of China (Grant No. 12404320).

\appendix
\section{Effects of lattice mismatches}\label{AppA}
\begin{figure}[H]
    \centering
    \includegraphics[width=1.0\linewidth]{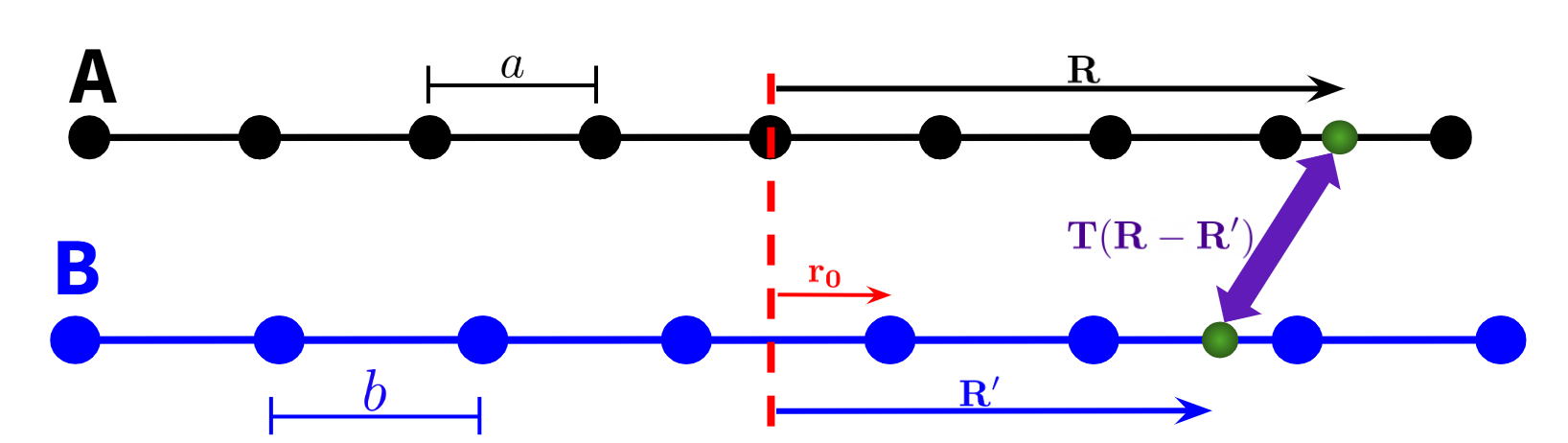}
    \caption{Tunneling of an electron(green ball) between two chains $A$ and $B$. One lattice site in chain $A$ is chosen as the origin and the second chain is shifted relatively by an amount $\mathbf r_{0}$. The amplitude for an electron at position $\mathbf{R'}$ in chain $B$ to tunnel to position $\mathbf{R}$ in chain $A$(and vice-versa) is $T(\mathbf{R}-\mathbf{R'})$.}
    \label{fig:shift_lattice}
\end{figure}

In this appendix, we will study how structural differences can affect the momentum of an electron tunneling between two lattices and explicitly illustrate these effects with a 1D toy-model. \par
We start with two lattices A and B, displaced with respect to each other by an amount $\mathbf{r}_{0}$, as illustrated in figure \ref{fig:shift_lattice}. We assume that the tunneling amplitude for an electron in position $\mathbf{R'}$ of lattice B to a position $\mathbf{R}$ of the lattice A depends on their relative separation, so that the tunneling Hamiltonian reads
\begin{align}
    \hat{H}_{t}= \int_{A} d\mathbf{R}\int_{B} d\mathbf{R'} T(\mathbf{R}-\mathbf{R'})\hat{\Psi}^{\dagger}_{A}(\mathbf{R})\hat{\Psi}_{B}(\mathbf{R'})  + {h.c}.
    \label{ec:real_tunnel}
\end{align}
We express the field operators for each lattice in terms of their respective Bloch states as:
\begin{align}
    \hat{\Psi}_{A}(\mathbf{R})=& \sum_{k,m}e^{i\mathbf{k}\cdot\mathbf{R}}u^{A}_{k,m}(\mathbf{R})\hat{a}_{\mathbf{k},m} \\
    \hat{\Psi}_{B}(\mathbf{R'})=& \sum_{p,n}e^{i\mathbf{p}\cdot(\mathbf{R'}-\mathbf{r}_{0})}u^{B}_{k,m}(\mathbf{R'}-\mathbf{r}_{0})\hat{b}_{\mathbf{p},m},
\end{align}
where $\mathbf{k}$($\mathbf{p}$) is a momentum inside the first Brillouin zone of A(B) and $m$($n$) is a band index. In this way, the tunneling Hamiltonian takes the form
\begin{align}
    \hat{H}_{t}=\sum_{\mathbf{k},\mathbf{p}}\sum_{m,n}T_{m,n}(\mathbf{k},\mathbf{p})\hat{a}^{\dagger}_{\mathbf{k},m}\hat{b}_{\mathbf{p},n}+{h.c}.
\end{align}
\begin{figure}
    \centering
    \includegraphics[width=1.0\linewidth]{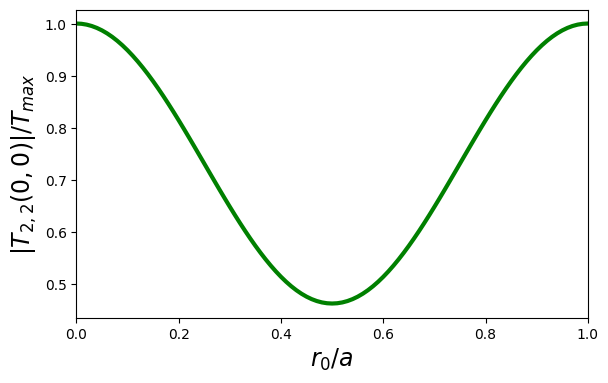}
    \caption{Tunneling amplitude as a function of the relative displacement $r_{0}$ between two identical chains of lattice constant $a$. Parameters used: $\{E_{0},V_{a},V_{b},a,b,L_{A},L_{B},\sigma\}=\{1,20,20,1,1,100,100,0.33\}$.  }
    \label{fig:constantshifts}
\end{figure}
with the tunneling amplitude in reciprocal space defined as
\begin{align}
    \nonumber T_{m,n}(\mathbf{k},\mathbf{p})&=\int_{A}d\mathbf{R}\int_{B}d\mathbf{R'}e^{i \mathbf{p}\cdot(\mathbf{R'}-\mathbf{r_{0}})}e^{-i\mathbf{k}\cdot \mathbf{R}}\\
    &\times(u^{A}_{\mathbf{k},m}(\mathbf{R}))^{*}u^{B}_{\mathbf{p},n}(\mathbf{R'}-\mathbf{r_{0}})T(\mathbf{R}-\mathbf{R'}).
    \label{ec:momentum-tunneling}
\end{align}
To illustrate the general behavior of the tunneling, we consider two one-dimensional chains modeled by sinusoidal lattice potentials for the electrons
\begin{align}
     \hat{H}_{0}^{\eta}=&\frac{\hat{p}^{2}}{2m}+2V_{\eta}\cos\left(\frac{2\pi}{\eta}x\right)
    \label{ec:chain_hamils}
\end{align}
with $\eta=a,b$. These Hamiltonians can be diagonalized using a plain-wave basis, so that its Bloch wave functions take the form
\begin{align}
    u^{\eta}_{k,m}(x)=\sum_{j}c^{\eta}_{m,j}(k)\frac{e^{i\cdot j\cdot G^{\eta}x}}{\sqrt{L_{\eta}}}
    \label{ec:Blochwave}
\end{align}
where $j$ is an integer, $L_{\eta}$ is the length of each chain, $G^{\eta}=2\pi/\eta$ and $c^{\eta}_{m,j}(k)$ are the coefficients obtained from the diagonalization of \ref{ec:chain_hamils} for each momentum $k$. Plugging equation \ref{ec:Blochwave} in \ref{ec:momentum-tunneling} and using a Gaussian ansatz for the tunneling $T(r-r')=t\exp\{-(r-r')^{2}/(2\sigma^{2})\}$ we obtain
\begin{widetext}    
\begin{align}
    \nonumber T_{m,n}(k,p) =& t\sum_{j,l}\frac{(c^{a}_{m,j}(k))^{*}c_{n,l}^{b}(p)}{\sqrt{L_{a}L_{b}}}\int_{-L_{a}/2}^{L_{a}/2}\int_{-L_{b}/2+r_{0}}^{L_{b}/2+r_{0}}drdr'e^{-i(k+j\cdot G^{a})r}e^{i(p+l \cdot G^{b})(r'-r_{0})}e^{-\frac{(r-r')^{2}}{2\sigma^{2}}}\\
     \approx & t\sigma \sqrt{\frac{2 \pi L_{a}}{L_{b}}}\sum_{j,l}(c^{a}_{m,j}(k))^{*}c_{n,l}^{b}(p)e^{-p^{2}\sigma^{2}/2}e^{-ir_{0}(p+l \cdot G^{b})}\text{sinc}\left( \frac{(p+l\cdot G^{b})-(k+j \cdot G^{b})}{2}L_{a}\right),
     \label{eq:1D_tunnel}
\end{align}
\end{widetext}
where we assume the chains are sufficiently long to obtain the explicit expression in the second line. 

We first consider a scenario where two identical chains were shifted relative to each other by $\mathbf r_{0}$. For numerical simulations we set $E_{0}=\hbar^{2}/(2ma^{2})$ as our unit of energy and examine the amplitude for electrons to tunnel from chain $A$ to chain $B$ while remaining in the same band(we chose $k=p=0$ and the second excited band for illustration purposes). In figure \ref{fig:constantshifts} we see that the effect of the shift is simply to reduce the overall amplitude of the tunneling, but since the systems have the same translational invariance then in principle the momentum is not greatly affected.

\begin{figure}[H]
    \centering
    \includegraphics[width=1.0\linewidth]{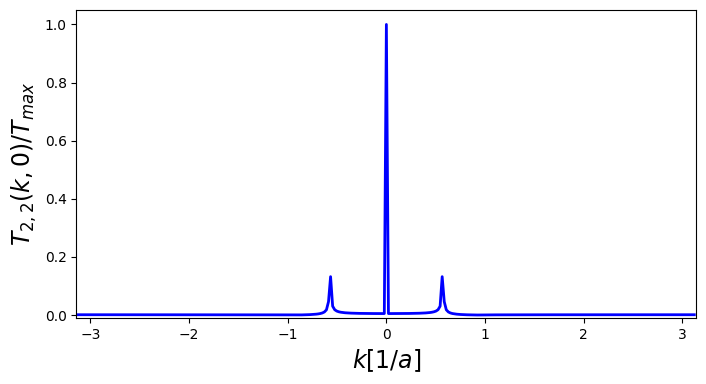}
    \caption{Tunneling amplitudes between chains with a small lattice mismatch. Parameters used are $\{E_{0},V_{a},V_{b},a,b,L_{A},L_{B},\sigma,r_{0}\}=\{1,20,16.5,1,1.1,300,330,0.33,0\}$.}
    \label{fig:smallmismatch}
\end{figure}
Next, we analyze a situation of small lattice mismatches by simulating two chains with lattice constants $a$ and $b=1.1a$ respectively. Figure \ref{fig:smallmismatch} shows the amplitude for electrons to tunnel from chain $A$ to chain $B$ at zero momentum. We see a dominant central  peak, indicating momentum conservation to a large degree. However, there are two small peaks separated by a distance $\Delta k=|2\pi/a-2\pi/b|$ from the center, which explicitly illustrate that momentum is conserved up to a difference of reciprocal lattice vectors.

\begin{figure}
    \centering
    \includegraphics[width=1.0\linewidth]{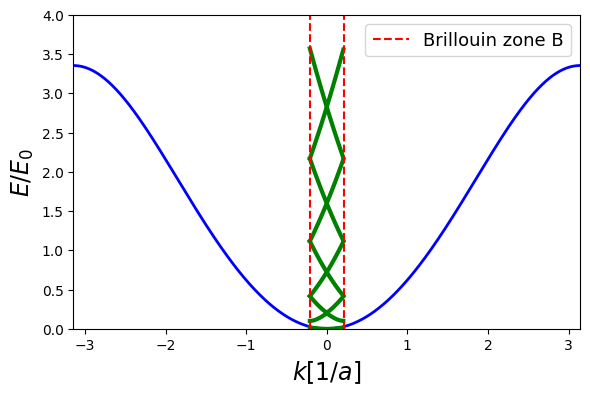}
    \includegraphics[width=1.0\linewidth]{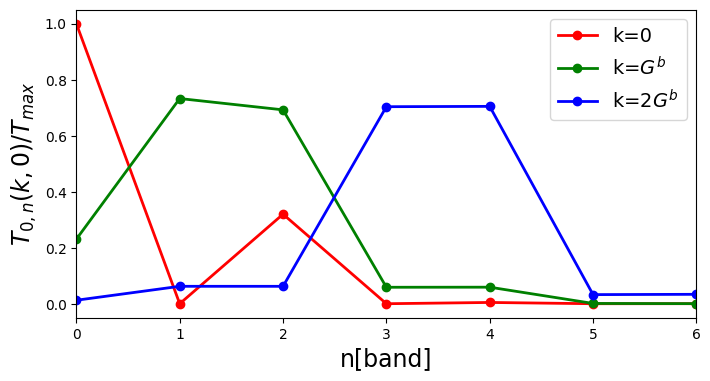}
    \caption{\textbf{Top:} Overlapped band structures of two chains with a large mismatch. \textbf{Bottom:} Amplitudes for an electron in chain $A$ with momentum $k$ to tunnel to the $n$-th band in chain $B$ with zero momentum. Parameters used are $\{E_{0},V_{a},V_{b},a,b,L_{A},L_{B},\sigma,r_{0}\}=\{1,10,0.04,1,15,1000,1500,0.33,0\}$.}
    \label{fig:moirenormal}
\end{figure}

Finally, we investigate a situation with large lattice mismatch. In figure \ref{fig:moirenormal} we set $a=1$, $b=15 a$ and plot the band structures of the chains in their respective first Brillouin zones. It is clear that some momentum values in $A$ can be far beyond the Brillouin zone boundaries of $B$. We then calculate the tunneling amplitudes from chain $A$ to chain $B$ for electrons with initial momenta that were multiples of the reciprocal lattice vector of chain B ($G^{b}=2\pi/b$). We conclude that high momentum electrons in lattice $A$ will tunnel to high bands(but at small momentum) in lattice $B$, therefore the model in the main manuscript would need to account for these effects if, for example, the lower material in figure \ref{Fig-setup} wasn't a moiré bilayer.

\section{Electron-exciton scattering matrix}\label{AppB}
Under the assumption that the low energy scattering in vacuum due to $V_{\text{ex}} (\mathbf{k})$ can be described by $s$-wave scattering with scattering amplitude $g$, one finds $\mathcal{T}$ is only an energy $\omega$ dependent function.
\begin{equation}
\mathcal{T}\left(\omega\right)=g+\frac{g}{N}\sum_{\mathbf{k}\in BZ}G_U\left(\mathbf{k},\omega\right)T\left(\omega\right)
\end{equation}
where BZ is the first Brillouin zone of the lattice. By solving the above equation, one has
\begin{equation}
\mathcal{T}\left(\omega\right)=\frac{1}{1/g-\frac{1}{N}\sum_{\mathbf{k}\in BZ}G_U\left(\mathbf{k},\omega\right)}.
\end{equation}
When $\omega=\delta+\epsilon_T$, the scattering is on resonance and $\mathcal{T}(\omega)$ diverges. Consequently, one obtains the expression for $g$ in terms of $\epsilon_T$
\begin{equation}
\frac{1}{g}=\frac{1}{N}\sum_{\mathbf{k}\in BZ}G_U\left(\mathbf{k},\delta+\epsilon_T\right)\approx\frac{1}{\epsilon_T}
\end{equation}
where $\epsilon_\mathbf{k}$ has been ignored in $G_U(\mathbf{k},\omega)$.
Therefore,
\begin{equation}\begin{aligned}
\mathcal{T}\left(\omega\right)\approx\frac{1}{1/\epsilon_T-\frac{1}{\omega-\delta}}
&=\frac{\epsilon_T\left(\omega-\delta\right)}{\omega-\delta-\epsilon_T}\\
&\xrightarrow[]{\omega\ll \delta}-\frac{\epsilon_T\delta}{\omega-\delta-\epsilon_T}\\
&\xrightarrow[]{\omega\ll \delta+\epsilon_T}\frac{\epsilon_T\delta}{\delta+\epsilon_T}.
\end{aligned}\end{equation}

\section{Second order energy shift}\label{AppC}
Since $G_U(\mathbf{k},\omega)$ is approximately independent of $\mathbf{k}$, $\sum_{\mathbf{k}}G_U(\mathbf{k},\omega)\exp(i\mathbf{k}\cdot \mathbf{r})\approx 0$ for $\mathbf r\neq0$. Thus one can ignore the contribution from the propagation between two excitons in upper layer to the energy shift and the second order perturbation theory gives
\begin{widetext}
\begin{equation}\begin{aligned}
E_2&=\sum_{n>0}\frac{|\langle \psi_n|\hat{V}_{\text {eff}}({\mathbf r_1})+\hat{V}_{\text {eff}}({\mathbf r_2})|\psi_0\rangle|^2}{E_{0}-E_n}\\
&=\kappa^2\sum_{n>0}\frac{\langle \psi_0|\left[\hat{n}_{\sigma_1}\left(\mathbf{r}_1\right)+\hat{n}_{\sigma_2}\left(\mathbf{r}_2\right)\right]|\psi_n\rangle\langle \psi_n|\left[\hat{n}_{\sigma_1}\left(\mathbf{r}_1\right)+\hat{n}_{\sigma_2}\left(\mathbf{r}_2\right)\right]|\psi_0\rangle}{E_{0}-E_n}\\
&=-\kappa^2\sum_{n>0}\frac{\langle \psi_0|\hat{n}_{\sigma_1}\left(\mathbf{r}_1\right)|\psi_n\rangle\langle \psi_n|\hat{n}_{\sigma_1}\left(\mathbf{r}_1\right)|\psi_0\rangle}{E_{n}-E_0}
-\kappa^2\sum_{n>0}\frac{\langle \psi_0|\hat{n}_{\sigma_2}\left(\mathbf{r}_2\right)|\psi_n\rangle\langle \psi_n|\hat{n}_{\sigma_2}\left(\mathbf{r}_2\right)|\psi_0\rangle}{E_{n}-E_0}\\
&\quad -\kappa^2\sum_{n>0}\frac{\langle \psi_0|\hat{n}_{\sigma_1}\left(\mathbf{r}_1\right)|\psi_n\rangle\langle \psi_n|\hat{n}_{\sigma_2}\left(\mathbf{r}_2\right)|\psi_0\rangle+\langle \psi_0|\hat{n}_{\sigma_2}\left(\mathbf{r}_2\right)|\psi_n\rangle\langle \psi_n|\hat{n}_{\sigma_1}\left(\mathbf{r}_1\right)|\psi_0\rangle}{E_{n}-E_0}
\end{aligned}\end{equation}
\end{widetext}
which leads to the final expression of Eq. \ref{SecondOrder} by applying the expression of the correlation function given in Ref. \cite{pitaevskii2016bose}.

\section{Spin-spin correlation for Eq. \ref{Eq-H}}\label{AppD}
Suppose the spin system is in the ferromagnetic phase with spin polarizing along $\mathbf{x}$-direction ($\phi\in [\pi/3, 2 \pi/3]$). One can apply the Holstein-Primakoff transformation
\begin{subequations}
\begin{equation}
\hat S_{x}\left(\mathbf{r}_i\right)=S-\hat{\alpha}_i^\dagger \hat{\alpha}_i
\end{equation}
\begin{equation}
\hat S^+\left(\mathbf{r}_i\right)=\hat S_{y}\left(\mathbf{r}_i\right)+i\hat S_{z}\left(\mathbf{r}_i\right)\approx \sqrt{2S}\hat{\alpha}_i
\label{Eq-HP-S+}\end{equation}
\begin{equation}
\hat S^-\left(\mathbf{r}_i\right)=\hat S_{y}\left(\mathbf{r}_i\right)-i\hat S_{z}\left(\mathbf{r}_i\right)\approx \sqrt{2S} \hat{\alpha}_i^\dagger
\label{Eq-HP-S-}\end{equation}
\label{Eq-HPT}\end{subequations}
where $S=1/2$ is the spin and $\alpha_i (\alpha_i^\dagger)$ is the bosonic annihilation (creation) operator at position $\mathbf{r}_i$. 
Within quadratic order in $\hat{\alpha}_i$ and $\hat{\alpha}_i^\dagger$ one has
\begin{equation}
\hat{H}_S=\hat{H}_0+\hat{H}_2
\label{Eq-HS-HPT}\end{equation}
where
\begin{equation}
\hat{H}_0=JS^2\sum_{\langle i,j\rangle}\cos\left(2\phi_{ij}\right)=3NJS^2\cos\left(2\phi\right)
\end{equation}
and
\begin{widetext}
\begin{equation}
\hat{H}_2=JS\sum_{\langle i,j\rangle}\left\{-2\cos\left(2\phi_{ij}\right)\hat{\alpha}_i^\dagger \hat{\alpha}_i+\left[\cos\left(2\phi_{ij}\right)+1\right]\hat{\alpha}_i^\dagger \hat{\alpha}_j+\frac{1}{2}\left[\cos\left(2\phi_{ij}\right)-1\right]\left(\hat{\alpha}_i\hat{\alpha}_j+\hat{\alpha}_i^\dagger \hat{\alpha}_j^\dagger\right)\right\}
\end{equation}
\end{widetext}
where $\phi_{ij}=-\phi_{ji}$ is applied and the linear terms in $\hat{\alpha}_i$ and $\hat{\alpha}_i^\dagger$ vanish automatically for ground state. 
To obtain the excitation spectrum, we Fourier transform $\hat{H}_2$ into the  momentum space using  $\hat{\alpha}_i=\left(1/\sqrt{N}\right)\sum_\mathbf{k}\alpha_\mathbf{k}e^{i\mathbf{k}\mathbf{r}_i}$ and find
\begin{widetext}
\begin{equation}
\hat{H}_2=JS\sum_\mathbf{k}\left[A_\mathbf{k}\alpha_\mathbf{k}^\dagger \alpha_\mathbf{k}-\frac{1}{2}B_\mathbf{k}\left(\alpha_{\mathbf{k}}\alpha_{-\mathbf{k}}+\alpha_{\mathbf{k}}^\dagger \alpha_{-\mathbf{k}}^\dagger\right)\right]
\label{Eq-H2-k}\end{equation}
where
\begin{subequations}
\begin{equation}\begin{aligned}
A_\mathbf{k}=-6\cos(2\phi)+
\left[\cos\left(2\phi\right)+1\right]\left[\cos(k_x)+2\cos(k_x/2)\cos(\sqrt{3}k_y/2)\right]
\end{aligned}\end{equation}
\begin{equation}
B_\mathbf{k}=-\left[\cos\left(2\phi\right)-1\right] \left[\cos k_x+2\cos \left(k_x/2\right) \cos \left(\sqrt{3}k_y/2\right)\right].
\end{equation}
\end{subequations}
\end{widetext}
With Bogoliubov approximation $\hat{\alpha}_\mathbf{k}=u_\mathbf{k}\hat{\beta}_\mathbf{k}+v_\mathbf{k}\hat{\beta}_{-\mathbf{k}}^\dagger$ and $u_\mathbf{k}^2-v_\mathbf{k}^2=1$ one finds $\hat{H}_2$ can be diagonalized by taking
\begin{subequations}
\begin{equation}
u_\mathbf{k}^2+v_\mathbf{k}^2=\frac{A_\mathbf{k}}{\sqrt{A_\mathbf{k}^2-B_\mathbf{k}^2}}
\end{equation}
\begin{equation}
2u_\mathbf{k}v_\mathbf{k}=\frac{B_\mathbf{k}}{\sqrt{A_\mathbf{k}^2-B_\mathbf{k}^2}},
\end{equation}
\end{subequations}
yielding the excitation spectrum
\begin{equation}
E_\mathbf{k}=JS\sqrt{A_\mathbf{k}^2-B_\mathbf{k}^2}.
\end{equation}
By applying the Fourier transform  $\hat S_{z}(\mathbf{k})=\left(1/\sqrt{N}\right)\sum_i \hat S_{z}(\mathbf{r}_i)\exp\left(-i \mathbf{k} \cdot \mathbf{r}_i\right)$, the static spin-spin correlation function in momentum space reads
\begin{equation}\begin{aligned}
\chi_{S_z,S_z}\left(\mathbf{k},0\right)&=\sum_{n>0}\frac{2|\langle \psi_n|S_{z}\left(\mathbf{k}\right)|\psi_0\rangle|^2}{E_{n}-E_{0}}\\
&=\sum_{n>0}\frac{2|\langle \psi_n|\frac{\sqrt{2S}}{2i}\left(\hat{\alpha}_\mathbf{k}-\hat{\alpha}_{-\mathbf{k}}^{\dagger} \right)|\psi_0\rangle|^2}{E_{n}-E_{0}}\\
&=S\sum_{n>0}\frac{|\langle \psi_n|\left(v_\mathbf{k}\hat{\beta}_{-\mathbf{k}}^\dagger-u_{\mathbf{k}}\hat{\beta}_{-\mathbf{k}}^{\dagger} \right)|\psi_0\rangle|^2}{E_{n}-E_{0}}\\
&=\frac{1}{J\left(A_\mathbf{k}+B_\mathbf{k}\right)}.
\end{aligned}\end{equation}
As a result, the spin-spin correlation in real space is given by
\begin{equation}\begin{aligned}
\langle\langle \hat S_z(\mathbf{r}_i),\hat S_z(\mathbf{r}_j)\rangle\rangle&=\frac{1}{N}\sum_\mathbf{k}\chi_{S_z,S_z}\left(\mathbf{k},0\right)e^{i\mathbf{k}\cdot \left(\mathbf{r}_i-\mathbf{r}_j\right)}\\
&=\frac{1}{NJ}\sum_\mathbf{k}\frac{e^{i\mathbf{k}\cdot\left(\mathbf{r}_i-\mathbf{r}_j\right)}}{A_\mathbf{k}+B_\mathbf{k}}.
\end{aligned}\label{Eq-chi-SzSz-Real}\end{equation}

If the system is in the $120^\circ$-1 ($120^\circ$-2) antiferromagnetic order, that is $\phi\in [2\pi/3, \pi]$ ($[0, \pi/3]$), one can first express the Hamiltonian in terms of the spin in the frame rotated around $\hat{z}-$axis by angle $\theta_i=\mathbf{Q}\cdot \mathbf{r}_i$ with $\mathbf{Q}=\left(-4\pi/3,0\right)$ ($\mathbf{Q}=\left(4\pi/3,0\right)$) and then treat the system in the rotated frame as the ferromagnetic order and apply the Holstein-Primakoff transformation shown in Eq. \ref{Eq-HPT} to calculate the spin-spin correlation function. Now we only perform explicit calculations for the $120^\circ$-1 antiferromagnetic order, as that for the  $120^\circ$-2 antiferromagnetic order is similar. The relationship between the spin in laboratory frame ($S_x, S_y, S_z$) and rotational frame ($S_x^R, S_y^R, S_z^R$) is
\begin{subequations}
\begin{equation}
\hat S_x=\hat S_x^R\cos \theta - \hat S_y^R\sin \theta
\end{equation}
\begin{equation}
\hat S_y= \hat S_x^R \sin \theta + \hat S_y^R \cos \theta
\end{equation}
\begin{equation}
\hat S_z=\hat S_z^R.
\end{equation}
\end{subequations}
Therefore, the Hamiltonian in the rotated frame is
\begin{widetext}
\begin{equation}\begin{aligned}
\hat H_S=J\sum_{\langle i,j\rangle}\bigg\{\hat S_{z}^R(\mathbf{r}_i)\hat S_{z}^R(\mathbf{r}_j)+&\cos \left(\theta_i-\theta_j+2\phi_{ij}\right)\left[\hat S_{x}^R(\mathbf{r}_i)\hat S_{x}^R(\mathbf{r}_j) +\hat S_{y}^R(\mathbf{r}_i) \hat S_{y}^R(\mathbf{r}_j)\right]\\
+&\sin \left(\theta_i-\theta_j+2\phi_{ij}\right)\left[\hat S_{x}^R(\mathbf{r}_i)\hat S_{y}^R(\mathbf{r}_j)- \hat S_{y}^R(\mathbf{r}_i)\hat S_{x}^R(\mathbf{r}_j) \right]\bigg\}.
\end{aligned}\end{equation}
\end{widetext}
Compared to Eq. \ref{Eq-H}, this Hamiltonian has only two differences: 1) $\hat S_{x,y,z}(\mathbf{r}_i)$ is replaced by $\hat S_{x,y,z}^R(\mathbf{r}_i)$, and 2) $2\phi_{ij}$ is replaced by $\theta_i-\theta_j+2\phi_{ij}$. Since $\theta_i-\theta_j=-(\theta_j-\theta_i)$, the dummy variable can be exchanged as in deriving Eq. \ref{Eq-HS-HPT}, and one can easily write down the Hamiltonian after performing the Holstein-Primakoff transformation given in Eq. \ref{Eq-HPT} (replacing $\hat S_{x,y,z}(\mathbf{r}_i)$ by $\hat S_{x,y,z}^R(\mathbf{r}_i)$). That is
\begin{equation}
\hat{H}_S=\hat{H}_0+\hat{H}_2
\end{equation}
where
\begin{equation}\begin{aligned}
\hat{H}_0&=JS^2\sum_{\langle i,j\rangle}\cos\left(\theta_i-\theta_j+2\phi_{ij}\right)\\
&=3NJS^2\cos\left(2\phi-\frac{2\pi}{3}\right)
\end{aligned}\end{equation}
and
\begin{widetext}
\begin{equation}
\hat{H}_2=JS\sum_{\langle i,j\rangle}\left\{-2\cos\left(\theta_i-\theta_j+2\phi_{ij}\right)\hat{\alpha}_i^\dagger \hat{\alpha}_i+\left[\cos\left(\theta_i-\theta_j+2\phi_{ij}\right)+1\right]\hat{\alpha}_i^\dagger \hat{\alpha}_j+\frac{1}{2}\left[\cos\left(\theta_i-\theta_j+2\phi_{ij}\right)-1\right]\left(\hat{\alpha}_i\hat{\alpha}_j+\hat{\alpha}_i^\dagger \hat{\alpha}_j^\dagger\right)\right\}.
\end{equation}
Fourier transform $H_2$ to momentum space, one obtains Eq. \ref{Eq-H2-k} with
\begin{subequations}
\begin{equation}
A_\mathbf{k}=-6\cos(2\phi-\frac{2\pi}{3})+\left[\cos\left(2\phi-\frac{2\pi}{3}\right)+1\right]\left[\cos(k_x)+2\cos(k_x/2)\cos(\sqrt{3}k_y/2)\right]
\end{equation}
and
\begin{equation}
B_\mathbf{k}=-\left[\cos\left(2\phi-\frac{2\pi}{3}\right)-1\right] \left[\cos k_x+2\cos \left(k_x/2\right) \cos \left(\sqrt{3}k_y/2\right)\right].
\end{equation}
\end{subequations}
\end{widetext}
These expressions can be obtained by  shifting $\phi$ by $\pi/3$ in those for the ferromagnetic order. Therefore, the spin-spin correlation function is same as that for the ferromagnetic order except for a shift of $\phi$ by $\pi/3$. Similarly, one can verify that the spin-spin correlation function for $120^\circ$-2 antiferromagnetic order is also same as that for ferromagnetic order except a shift of $\phi$ by $-\pi/3$.

\section{Spin correlations for a superconducting sample}\label{AppE}
In this appendix we derive the correlation functions for a Hamiltonian which is more general than the one shown in the main text. Following generalized BCS theory, it is well known that the single-band BdG Hamiltonian can be diagonalized as 
\begin{align}
     \nonumber \hat{H}=&\frac{1}{2}\sum_{\mathbf{k}}\hat{\Psi}_{k}^{\dagger} \begin{pmatrix}
         \xi_{\mathbf{k}}\hat{\sigma}_{0} \quad \hat{\Delta}_{\mathbf{k}} \\  \hat{\Delta}_{\mathbf{k}}^{\dagger} \quad  -\xi_{\mathbf{k}}\hat{\sigma}_{0}
     \end{pmatrix} \hat{\Psi}_{\mathbf{k}} + const\\
     =&\sum_{\mathbf{k}}E_{\mathbf{k}}(\gamma^{\dagger}_{\mathbf{k}\uparrow} \gamma_{\mathbf{k}\uparrow}+\gamma^{\dagger}_{\mathbf{k}\downarrow} \gamma_{\mathbf{k}\downarrow}).
     \label{ec:BdG}
 \end{align}
Here $\hat{\Psi}_{\mathbf{k}}:=(\hat{c}_{\mathbf{k}\uparrow},\hat{c}_{\mathbf{k}\downarrow},\hat{c}^{\dagger}_{-\mathbf{k}\uparrow},\hat{c}^{\dagger}_{-\mathbf{k}\downarrow})^{T}$ is the Nambu vector, $\hat{\Delta}_{\mathbf{k}}$ is the gap function matrix,(which includes pairing amplitudes between electrons with both parallel and anti-parallel spins), and $\hat{\Gamma}_{\mathbf{k}}=(\gamma_{\mathbf{k}\uparrow},\gamma_{\mathbf{k}\downarrow},\gamma^{\dagger}_{-\mathbf{k}\uparrow}, \gamma^{\dagger}_{-\mathbf{k}\downarrow})$ are the Bogoliubov quasi-particle operators, which are related to the electronic operators by the unitary transformation
\begin{align}
    \hat{\Psi}_{\mathbf{k}}=\hat{U}_{\mathbf{k}}\hat{\Gamma}_{\mathbf{k}}, \quad \hat{U}_{\mathbf{k}} := \begin{pmatrix}
        \hat{u}_{\mathbf{k}} \quad \hat{v}_{\mathbf{k}} \\
        \hat{v}^{*}_{\mathbf{-k}} \quad \hat{u}^{*}_{\mathbf{-k}}
    \end{pmatrix}\quad.
    \label{ec:transforms1}
\end{align}
For unitary pairing, the transformation takes a simplified form \cite{general_supercond}
\begin{align}
    \hat{u}_{\mathbf{k}}=\frac{E_{\mathbf{k}}+\xi_{\mathbf{k}}}{\sqrt{2E_{\mathbf{k}}(E_{\mathbf{k}}+\xi_{\mathbf{k}})}}\hat{\sigma}_{0}, \quad \hat{v}_{\mathbf{k}}=\frac{-1}{\sqrt{2E_{\mathbf{k}}(E_{\mathbf{k}}+\xi_{\mathbf{k}})}}\hat{\Delta}_{\mathbf{k}},
    \label{ec:transforms2}
\end{align}
where $E_{\mathbf{k}}=\sqrt{\xi_{\mathbf{k}}^{2}+|\Delta_{\mathbf{k}}|^{2}}$. Using the transformations \ref{ec:transforms1} we can express the operators in equation \ref{Veff} in terms of the quasi-particle operators as
\begin{align}
    \nonumber \hat{c}^{\dagger}_{\mathbf{p}\sigma}\hat{c}_{\mathbf{q}\sigma}= &u_{\mathbf{p}}u_{\mathbf{q}}\gamma_{\mathbf{p}\sigma}^{\dagger}\gamma_{\mathbf{q}\sigma}+v^{*}_{\mathbf{p};\sigma\sigma}v_{q;\sigma \sigma}\gamma_{-\mathbf{p}\sigma}\gamma_{-\mathbf{q}\sigma}^{\dagger}\\
    \nonumber+&v^{*}_{\mathbf{p};\sigma\sigma}v_{q;\sigma-\sigma}\gamma_{-\mathbf{p}\sigma}\gamma^{\dagger}_{-\mathbf{q}-\sigma}+v^{*}_{\mathbf{p};\sigma-\sigma}v_{\mathbf{q};\sigma \sigma}\gamma_{-\mathbf{p}-\sigma}\gamma^{\dagger}_{-\mathbf{q}\sigma} \\
    \nonumber+&v_{\mathbf{p};\sigma-\sigma}^{*}v_{\mathbf{q};\sigma-\sigma}\gamma_{-\mathbf{p}-\sigma}\gamma^{\dagger}_{-\mathbf{q}-\sigma}-u_{\mathbf{p}}v_{\mathbf{q};\sigma \sigma}\gamma^{\dagger}_{\mathbf{p}\sigma}\gamma^{\dagger}_{-\mathbf{q}\sigma}\\
    \nonumber-&u_{\mathbf{p}}v_{\mathbf{q};\sigma-\sigma}\gamma^{\dagger}_{\mathbf{p}\sigma}\gamma^{\dagger}_{-\mathbf{q}-\sigma}-v^{*}_{\mathbf{p};\sigma\sigma}u_{\mathbf{q}}\gamma_{-\mathbf{p}\sigma}\gamma_{\mathbf{q}\sigma}\\
    -&v^{*}_{\mathbf{p};\sigma -\sigma}u_{\mathbf{q}}\gamma_{-\mathbf{p}-\sigma}\gamma_{\mathbf{q}\sigma}.
    \label{ec:long_bogol}
\end{align}
The previous expression allows us to explicitly compute the matrix elements of density operators $\hat{n}_{\mathbf{r},\sigma}$. We now assume that the superconducting ground-state has no Bogoliubov quasi-particles so that  $\gamma_{\mathbf{k}\sigma}|\psi_0\rangle=0$ and $\langle \psi_0 |\gamma_{\mathbf{k}\sigma}\gamma^{\dagger}_{\mathbf{k'}\sigma'}|\psi_0\rangle=\delta_{\mathbf{k},\mathbf{k'}}\delta_{\sigma, \sigma'}$. From the previous analysis, we find that the only excited states that will yield non-vanishing matrix elements $\langle \psi_0|\hat{n}_{\mathbf{r}\sigma}|\psi \rangle$ are those of the form $|\psi\rangle =|\mathbf{k},\sigma;\mathbf{k'},\sigma'\rangle:=\gamma^{\dagger}_{\mathbf{k}\sigma}\gamma^{\dagger}_{\mathbf{k'}\sigma'}|\psi_0\rangle$, therefore we obtain

\begin{equation*}
\begin{cases} 
    \langle \psi_0|\hat{n}_{\mathbf{r},\downarrow}|\mathbf{k},\downarrow;\mathbf{k'},\uparrow\rangle&= -\frac{1}{N}e^{i(\mathbf{k+k'})\cdot \mathbf{r}} v^{*}_{\mathbf{-k'};\downarrow \uparrow} u_{\mathbf{k}}\\
    \langle \psi_0 | \hat{n}_{\mathbf{r},\downarrow} |\mathbf{k},\downarrow;\mathbf{k'},\downarrow\rangle&=\frac{-1}{N} e^{i(\mathbf{k+k'})\cdot \mathbf{r}}(u_{\mathbf{k}}v^{*}_{\mathbf{-k'},\downarrow \downarrow}-u_{\mathbf{k'}}v^{*}_{\mathbf{-k},\downarrow \downarrow}) \\  
    \langle \psi_0 |\hat{n}_{\mathbf{r},\uparrow}|\mathbf{k},\downarrow;\mathbf{k'},\uparrow \rangle&=\frac{1}{N}e^{i(\mathbf{k+k'})\cdot \mathbf{r}}v^{*}_{\mathbf{-k};\uparrow \downarrow}u_{\mathbf{k'}} \\ 
    \langle \psi_0 | \hat{n}_{\mathbf{r},\downarrow} |\mathbf{k},\uparrow;\mathbf{k'},\uparrow\rangle&= 0\\
    \langle \psi_0 | \hat{n}_{\mathbf{r},\uparrow} |\mathbf{k},\downarrow;\mathbf{k'},\downarrow\rangle&= 0.
\end{cases}
    \label{ec:matel2}
\end{equation*}

Since the excited states are pairs of Bogoliubons with different momenta and spins, equation \ref{SecondOrder} for a superconductor turns into a double sum over $\mathbf{k}$ and $\mathbf{k'}$. Starting with the parallel-spin correlation, we need to be careful with double counting over some states, because $|\mathbf{k},\downarrow;\mathbf{k'},\downarrow\rangle=-|\mathbf{k'},\downarrow;\mathbf{k},\downarrow\rangle$. The is easily resolved by adding a factor of $1/2$ wherever necessary, that is
\begin{widetext}    
\begin{align}
    \nonumber &\langle\langle \hat n_{\downarrow}(\mathbf{r}),\hat n_{\downarrow}(\mathbf{0})\rangle\rangle\\
    \nonumber=&\frac{1}{2}\sum_{\mathbf{k},\mathbf{k'}}\frac{\langle \mathbf{k},\downarrow;\mathbf{k'},\downarrow |\hat{n}_{\mathbf{r},\downarrow}|\psi_0\rangle \langle \psi_0 | \hat{n}_{\mathbf{0},\downarrow}|\mathbf{k},\downarrow;\mathbf{k'},\downarrow \rangle+c.c}{E_{\mathbf{k}}+E_{\mathbf{k'}}} 
    + \sum_{\mathbf{\mathbf{k}},\mathbf{\mathbf{k'}}}\frac{\langle \mathbf{k},\downarrow;\mathbf{k'},\uparrow |\hat{n}_{\mathbf{r},\downarrow}|\psi_0\rangle \langle \psi_0 | \hat{n}_{\mathbf{0},\downarrow}|\mathbf{k},\downarrow;\mathbf{k'},\uparrow \rangle+c.c}{E_{\mathbf{k}}+E_{\mathbf{k'}}} \\
    \nonumber=&\frac{1}{2N^{2}}\sum_{\mathbf{\mathbf{k}},\mathbf{\mathbf{k'}}}\frac{e^{-i(\mathbf{k+k'})\cdot \mathbf{r}}(u^{*}_{\mathbf{k}}v_{\mathbf{-k'};\downarrow \downarrow}-u^{*}_{\mathbf{k'}}v_{\mathbf{-k};\downarrow \downarrow})(u_{\mathbf{k}}v^{*}_{\mathbf{-k'};\downarrow \downarrow}-u_{\mathbf{k'}}v^{*}_{\mathbf{-k};\downarrow \downarrow})+ c.c }{E_{\mathbf{k}}+E_{\mathbf{k'}}} 
    +\frac{1}{N^{2}}\sum_{\mathbf{k},\mathbf{k'}}\frac{e^{-i(\mathbf{k+k'})\cdot \mathbf{r}}|u_{\mathbf{k}}|^{2}|v_{\mathbf{-k'};\downarrow \uparrow}|^{2}+c.c}{E_{\mathbf{k}}+E_{\mathbf{k'}}} \\
    = &\frac{2}{N^{2}}\sum_{\mathbf{\mathbf{k}},\mathbf{\mathbf{k'}}}\frac{|u_{\mathbf{k}}|^{2}(|v_{\mathbf{-k'};\downarrow \uparrow}|^{2}+|v_{\mathbf{k'};\downarrow \downarrow}|^{2})}{E_{\mathbf{k}}+E_{\mathbf{k'}}}\cos((\mathbf{k+k'})\cdot \mathbf{r}) 
    -\frac{2}{N^{2}}\sum_{\mathbf{\mathbf{k}},\mathbf{\mathbf{k'}}}\frac{(u_{\mathbf{k}}v_{\mathbf{-k};\downarrow \downarrow})^{*}(u_{\mathbf{k'}}v_{\mathbf{-k'};\downarrow \downarrow})e^{-i(\mathbf{k+k'})\cdot \mathbf{r}}}{E_{\mathbf{k}}+E_{\mathbf{k'}}}.
\label{ec:almost_corr_parall_spin}
\end{align}
\end{widetext}
In the last step, we used summation over dummy indices to group similar terms. The expression can be further simplified if we recall the explicit form of the transformations \ref{ec:transforms2} and see that
\begin{align}
    \nonumber u_{\mathbf{k}}v_{\mathbf{-k};\sigma\sigma'}&=\left( \frac{E_{\mathbf{k}}+\xi_{\mathbf{k}}}{\sqrt{2E_{\mathbf{k}}(E_{\mathbf{k}}+\xi_{\mathbf{k}})}}\right)\left( \frac{-\Delta_{\mathbf{-k};\sigma\sigma'}}{\sqrt{2E_{\mathbf{k}}(E_{\mathbf{k}}+\xi_{\mathbf{k}})}}\right)\\
    &=-\frac{\Delta_{-\mathbf{k},\sigma\sigma'}}{2E_{\mathbf{k}}}.
    \label{ec:usefulident}
\end{align}
Plugging \ref{ec:usefulident} in \ref{ec:almost_corr_parall_spin}, and using the anti-symmetry property of the gap function matrix $\hat{\Delta}_{\mathbf{k}}=-\hat{\Delta}_{-\mathbf{k}}$, the parallel spin correlation takes the form
\begin{widetext}

\begin{align}
    \nonumber \langle\langle \hat n_{\downarrow}(\mathbf{r}),\hat n_{\downarrow}(\mathbf{0})\rangle\rangle=&\frac{2}{N^{2}}\sum_{\mathbf{k,k'}}\frac{|u_{\mathbf{k}}|^{2}|v_{\mathbf{k'}}|^{2}}{E_{\mathbf{k}}+E_{\mathbf{k'}}}\cos((\mathbf{k+k'})\cdot \mathbf{r})
    +\frac{1}{2N^{2}}\sum_{\mathbf{k,k'}}\frac{1}{E_{\mathbf{k}}+E_{\mathbf{k'}}}\frac{ e^{-i(\mathbf{k+k'})\cdot \mathbf{r}}}{E_{\mathbf{k}}E_{\mathbf{k'}}}\Delta_{\mathbf{k};\downarrow \downarrow}^{*} \Delta_{\mathbf{-k'},\downarrow \downarrow}.
\end{align}
\end{widetext}
For the antiparallel spin correlation $\langle\langle \hat n_{\downarrow}(\mathbf{r}),\hat n_{\uparrow}(\mathbf{0})\rangle\rangle$, the matrix elements involving pairs of Bogoliubons with the same spin yield no contributions, therefore we sum over the states with pairs Bogoliubons of opposite spins. Using this fact, and recalling equation \ref{ec:usefulident} and the anti-symmetry of the cooper-pair matrix, we get
\begin{widetext}    
\begin{align}
    \nonumber \langle\langle \hat n_{\downarrow}(\mathbf{r}),\hat n_{\uparrow}(\mathbf{0})\rangle\rangle=&\sum_{\mathbf{k},\mathbf{k'}}\frac{1}{E_{\mathbf{k}}+E_{\mathbf{k'}}}\left( \langle \mathbf{k}, \downarrow ; \mathbf{k'} \uparrow|\hat{n}_{\mathbf{r},\uparrow}|\psi_0 \rangle \langle \psi_0 |\hat{n}_{\mathbf{0},\downarrow}|\mathbf{k}, \downarrow ; \mathbf{k'} \uparrow \rangle + c.c\right) \\
    \nonumber =&\frac{-1}{N^{2}}\sum_{\mathbf{k},\mathbf{k'}}\frac{1}{E_{\mathbf{k}}+E_{\mathbf{k'}}}\left( e^{-i(\mathbf{k+k'})\cdot \mathbf{r}}  v_{\mathbf{-k},\uparrow \downarrow}u_{\mathbf{k}} v^{*}_{\mathbf{-k'},\downarrow\uparrow}u_{\mathbf{k'}}+c.c\right)\\
    \nonumber =& -\frac{1}{4N^{2}}\sum_{\mathbf{k,k'}}\frac{1}{E_{\mathbf{k}}+E_{\mathbf{k'}}}\frac{1}{E_{\mathbf{k}}E_{\mathbf{k'}}}\left(e^{-i(\mathbf{k+k'})\cdot \mathbf{r}}\Delta_{\mathbf{-k},\uparrow\downarrow}\Delta^{*}_{\mathbf{-k'},\downarrow \uparrow} + e^{i(\mathbf{k+k'})\cdot \mathbf{r}}\Delta_{\mathbf{-k},\uparrow\downarrow}^{*}\Delta_{\mathbf{-k'},\downarrow \uparrow} \right) \\
    \nonumber =& \frac{1}{4N^{2}}\sum_{\mathbf{k,k'}}\frac{1}{E_{\mathbf{k}}+E_{\mathbf{k'}}}\frac{e^{-i(\mathbf{k+k'})\cdot \mathbf{r}}}{E_{\mathbf{k}}E_{\mathbf{k'}}}\Delta_{\mathbf{-k},\uparrow\downarrow}\Delta^{*}_{\mathbf{k'},\uparrow \downarrow}+\frac{1}{4N^{2}}\sum_{\mathbf{k,k'}}\frac{1}{E_{\mathbf{k}}+E_{\mathbf{k'}}}\frac{e^{-i(\mathbf{k+k'})\cdot \mathbf{r}}}{E_{\mathbf{k}}E_{\mathbf{k'}}}\Delta^{*}_{\mathbf{k},\uparrow\downarrow}\Delta_{\mathbf{-k'},\uparrow \downarrow}  \\
    =&\frac{1}{2N^{2}}\sum_{\mathbf{k,k'}}\frac{1}{E_{\mathbf{k}}+E_{\mathbf{k'}}}\frac{e^{-i(\mathbf{k+k'})\cdot \mathbf{r}}}{E_{\mathbf{k}}E_{\mathbf{k'}}}\Delta_{\mathbf{-k},\uparrow\downarrow}\Delta^{*}_{\mathbf{k'},\uparrow \downarrow}.
    \label{ec:ec:corr_antiparall_spin}
\end{align}
\end{widetext}
This is the correlation function studied in section \ref{section:Supercond}.

\bibliography{dingshsh-ref}

\end{document}